\documentclass[10pt]{article}

\usepackage{amsmath}
\usepackage{amssymb}

\usepackage{graphicx}

\usepackage{cite}

\usepackage{color} 

\usepackage[labelfont=bf,labelsep=period,justification=raggedright]{caption}

\makeatletter
\renewcommand{\@biblabel}[1]{\quad#1.}
\makeatother

\makeatletter
\renewcommand{\@biblabel}[1]{\quad#1.}
\makeatother

\date{}

\newcommand{\bea}{\begin{eqnarray} \nonumber }
\newcommand{\eea}{\end{eqnarray}}
\newcommand{\bi}{\begin{itemize}}
\newcommand{\ei}{\end{itemize}}

\newcommand{\s}{S}
\newcommand{\smoy}{\overline{\s}}

\newcommand{\sbvot}{S_{PS}}
\newcommand{\swc}{S_{b\equiv c}}
\newcommand{\swa}{S_{b\equiv a}}
\newcommand{\nv}{N_v}
\newcommand{\na}{N_a}
\newcommand{\nc}{N_c}
\newcommand{\nn}{N_{bn}}
\newcommand{\nnu}{N_{n}}
\newcommand{\nwh}{N_{b}}
\newcommand{\pv}{p_v}
\newcommand{\pa}{p_a}
\newcommand{\pc}{p_c}
\newcommand{\pn}{p_{bn}}
\newcommand{\pnu}{p_{n}}
\newcommand{\pwh}{p_{b}}
\newcommand{\paai}{p_{a,\,\alpha_i}}
\newcommand{\pcai}{p_{c,\,\alpha_i}}
\newcommand{\pnai}{p_{bn,\,\alpha_i}}
\newcommand{\paa}{p_{a,\,\alpha}}
\newcommand{\pca}{p_{c,\,\alpha}}
\newcommand{\pna}{p_{bn,\,\alpha}}
\newcommand{\sa}{S_{\alpha}}
\newcommand{\sai}{S_{\alpha_i}}
\newcommand{\ta}{\tau_{3,\,\alpha}}
\newcommand{\tai}{\tau_{3,\,\alpha_i}}

\begin{document}
\begin{flushleft}
{\Large
\textbf{Between order and disorder:\\a `weak law' on recent electoral behavior among urban voters?}
}
\\
Christian Borghesi$^{1,2 \ast}$, 
Jean Chiche$^{3}$, 
Jean-Pierre Nadal$^{1,4}$
\\
\bf {1} Centre d'Analyse et de Math\'ematique Sociales (CAMS), Centre National de la Recherche Scientifique \& Ecole des Hautes Etudes en Sciences Sociales, Paris, France
\\
\bf{2} Laboratoire de Physique Th\'eorique et Mod\'elisation (LPTM), Centre National de la Recherche Scientifique \& Universit\'e de Cergy-Pontoise, France
\\
\bf {3} Centre de recherches politiques de Sciences Po (CEVIPOF), Centre National de la Recherche Scientifique \& Sciences Po, Paris, France
\\
\bf {4} Laboratoire de Physique Statistique (LPS), Centre National de la Recherche Scientifique / \'Ecole Normale Sup\'erieure / Universit\'e Pierre et Marie Curie / Universit\'e Paris Diderot, Paris, France
\\
$\ast$ E-mail: borghesi@msh-paris.fr
\end{flushleft}

\section*{Abstract}
A new viewpoint on electoral involvement is proposed from the study of the statistics of the proportions of abstentionists, blank and null, and votes according to list of choices, in a large number of national elections in different countries. Considering 11 countries without compulsory voting (Austria, Canada, Czech Republic, France, Germany, Italy, Mexico, Poland, Romania, Spain and Switzerland), a stylized fact emerges for the most populated cities when one computes the entropy associated to the three ratios, which we call the entropy of civic involvement of the electorate. The distribution of this entropy (over all elections and countries) appears to be sharply peaked near a common value. This almost common value is typically shared since the 1970's by electorates of the most populated municipalities, and this despite the wide disparities between voting systems and types of elections. Performing different statistical analyses, we notably show that this stylized fact reveals particular correlations between the blank/null votes and abstentionists ratios.

We suggest that the existence of this hidden regularity, which we propose to coin as a `weak law on recent electoral behavior among urban voters', reveals an emerging collective behavioral norm characteristic of urban citizen voting behavior in modern democracies. Analyzing exceptions to the rule provide insights into the conditions under which this normative behavior can be expected to occur.



\section*{Introduction}

Each election yields a variable proportion of citizens not taking part in the vote.
The proportion of the uninvolved population -- either by non-registering, abstaining or voting blank or null -- has been much less studied than the vote itself.

Nowadays such behaviors are increasing among the longest-established democracies and their meaning may be changing. Besides passive abstention (due to carelessness or indifference), an active refusal of vote -- possibly bearing a political message -- is rising among population categories which are usually taking part in the election.

\subsubsection*{The modalities of withdrawal~\cite{Mayer-2010}}
To measure this phenomenon accurately, we first need to define the non-voter turnout. The boundary between voters and non-voters is indeed blurred as several intermediate behaviors exist, such as non-registering or blank vote.

The potential voter population depends on the legal requirements of citizenship, residency and capacity. Registration on the electoral roll does not necessarily imply voting. Moreover, the diversity of enumeration methods from one country to another makes it difficult to compare directly ratios of voters. The main trend consists in comparing abstention to the number of citizens entitled to vote (VEP: Voting Eligible Population). However, in the United States for instance, abstention was calculated until recently by comparison to the population above the voting age, including foreigners (VAP: Voting Age Population), the corresponding abstention rate often reaching 50\%. Another bias stems from the fact that some countries made voting compulsory (namely Belgium, Luxembourg, Greece, and for a time the Netherlands, Austria and Italy). Without compulsory voting, a declining voter turnout is observed since the 1980s in established democracies.

Moreover, the meaning of blank and null vote is not obvious. They could be considered at first sight as equivalent to abstention or non-registering, since they seem to translate an absence of choice. This hypothesis would be in agreement with the systematic reviews of the minutes of polling stations for instance.

Abstention has been primarily considered to be a micro-level phenomenon. But is it really? Several studies have proven that socio-economic characteristics such as gender~\cite{Dalton-1996,Topf}, age~\cite{Franklin-1996}, education~\cite{Verba-1995,Franklin-book} and ethnicity~\cite{Cees-Franklin-1996} have an influence on electoral non-participation. To what extent does living in a community with low level of electoral involvement influence a voter?

\subsubsection*{The political and institutional context of the election}
The comparative database collected by the Institute for Democracy and Electoral Assistance (IDEA~\cite{idea}) gathered data from elections in 171 countries from 1945 to 1999. It shows that participation rates are slightly higher in countries that have adopted a system of proportional representation, offering a larger choice to voters than those which have a majority or mixed systems. The highest turnout recorded (over 83\% observed in both Malta and Ireland) corresponds to the system of `single transferable vote' which gives the voter a large liberty margin. (This system, called Hare system of voting, is a variant of proportional representation where the voters rank the candidates according to their preferences.)

The nature of the election may be important too, depending on the context. In France for example, as the president has a lot of power, the participation rate of the presidential election is especially high when compared to the parliamentary election.

\subsubsection*{Abstention and Blank and null votes}
The reason why analysis of political sciences are paying little attention to blank and null votes is mostly based on the fact that these ballots are representing a very small number. Typically, these votes are aggregated within a single category, Blank and null votes, in some countries simply called Null (or Invalid) votes. Multitudinous studies have demonstrated from the 1950s on that null ballots were subdivided at random, according to the law of large number and distributed haphazardly for a given manner of voting~\cite{Lancelot}. The analysis of each voting office is still confirming that. However, the blank votes are more sensible to the conjuncture of consultation and are taking, with regard to abstention, a more complex signification.

Statistical analysis shows an often quite important negative correlation between abstention and Blank or invalid votes. In France, notably, it has often been observed that the more rural the municipality, the larger the ratio of Blank and null ballots. By contrast, the more populated the city, the larger the abstention ratio. However, the link between Abstention and Blank and null ballots becomes more complex in urban context. The urbanization has led to important changes in lifestyle and therefore in the voting behavior in large municipalities. Voters casting a blank vote are having motivations closer to voters abstaining for political reasons. This ``civic abstention'', as Alain Lancelot called it, expresses a particular attitude regarding the voting procedure~\cite{Lancelot,Zulfikarpasic}. This political attitude of ``withdrawal'' or political ``offside'' is not easy to analyze.

\subsubsection*{Looking for stylized facts}
In this paper, we analyze electoral data in order to better understand the interrelation between Abstention, Blank and null ballots and the expression of the vote, focusing on highly populated municipalities and recent elections. For this aim, we consider together the three values: Abstention, Blank/null and Valid votes ratios. We identify statistical regularities with an approach in the spirit of recent statistical physics analysis of elections data -- see e.g. \cite{costa_filho_scaling_vot,lyra_bresil_el,gonzalez_bresil_inde_el,fortunato_universality,daisy-model,araripe_role_parties,growth_model_vote,araujo_tactical_voting,these,diffusive_field,universality_candidates,turnout-stat,thurner-irregularities}.

By analyzing a large number of elections in 11 different countries without compulsory voting, we point out that they share a common feature when considering highly populated municipalities in recent elections (as specified later). Introducing a measure of civic involvement of electorate, we show that this quantity exhibits a sharply peaked distribution around a common value. Moreover we suggest that this common stylized fact, that we propose to call a `weak law on recent electoral behavior among urban voters', reveals an emerging collective behavioral norm, typical of citizen voting behavior in modern democracies.

The paper is organized as follows. First we describe the dataset used in this study, at three different scales (at the municipality scale, at larger scale but for older times, and at the polling station level when it is possible). Then, we introduce and discuss what we call the involvement entropy. We then analysis electoral data according to this measure, and give signs of existence of a possible norm revealed by a common-value of this measure. The Appendix S1 in the Supporting Information (SI) gives more details when it is necessary.

\section*{Materials and Methods}

\subsection*{Dataset}

In this paper we analyze electoral data at three different scales. (1) Data aggregated at the municipality scale. By this way, we study phenomena with respect to the population size of municipalities. The 76 elections studied in this paper at municipality level are mostly recent, after 1990, and are taken from 11 different countries (Austria, Canada, Czech Republic, France, Germany, Italy, Mexico, Poland, Romania, Spain and Switzerland). (2) Electoral data aggregated at large scale, e.g. national, provincial, etc. Here, we focus the analysis on time evolution. Countries studied for their historical aspects are those which are studied at the municipality scale. The study begins at the earliest year as possible, i.e. at the beginning of so-called democratic regimes, after World War II, and even earlier for some cases (e.g. 1884 for the $\approx 530$ Swiss referendums). (3) Electoral data aggregated at the polling station level. Polling stations over the 100 most populated municipalities are analyzed, whenever it is possible to do so (i.e. for Canada, France, Mexico, Poland and Romania). Some intra-towns phenomena are investigated by this way.

Some elections are studied as a function of the number $N$ of registered voters by municipality. This is the case when the following conditions are valid: (1) elections in a democratic country with no compulsory voting, and no duty against people who do not vote; (2) the number of registered voters by municipality is well established (in particular this excludes from our study both the U.S.A. and England); (3) available data provide for each municipality, at least, the number of registered voters, the number of votes or the turnout rate, and the number of valid votes. We note that all countries for which we have the data at the municipality scale have more than 2000 municipalities, which allows us to make statistical analysis. Moreover, all elections studied here are national ones, except for \textit{Land} Parliament elections in Germany. Lastly, the choice of the studied elections is not rooted on a plan but simply on the availability of electoral data.

Among these 76 elections, 31 of them are also analyzed at polling station level in the 100 most populated town: 5 from Canada ($\approx 25000$ polling stations), 13 from France ($\approx 7000$ polling stations), 4 from Mexico ($\approx 55000$ ballot boxes), 11 from Poland ($\approx 8000$ polling stations), and 4 from Romania ($\approx 6000$ polling stations). Tab.\ref{tdataset} summarizes the set of elections studied in this paper, and more details on these data are given in Appendix S1, Section~A.

The Appendix S1, Section~A, gives more information about the set of (public) electoral data studied in this paper. Most of them can be directly downloaded from official websites (see References in Appendix S1). Part of the database used in this paper can also be directly downloaded from~\cite{free-data}.

\subsection*{Abstentions, valid votes and blank or null votes} 
\begin{figure}[h!]
  \centering
  \includegraphics[width=0.45\linewidth, height=4cm, clip=true]{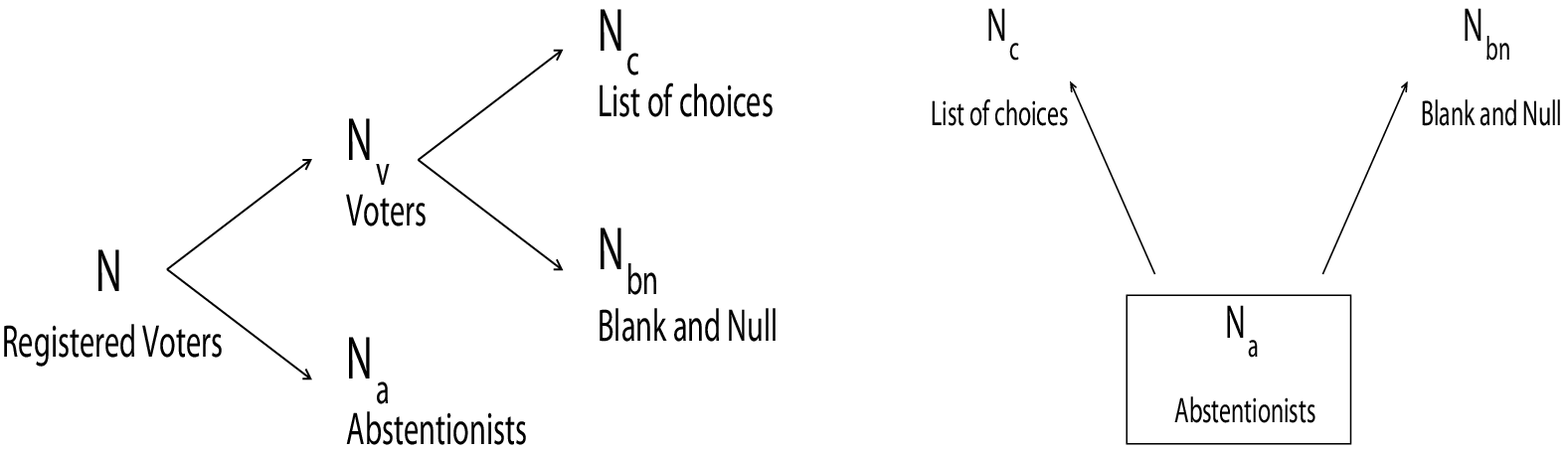}
  \caption{\bf{Electoral mobilization: categorization of registered voters according to their voting behavior.} The latter may result from
either two sequential binary decisions (left), or two mutually exclusive binary decisions (right), or (see Appendix S1, Section D.2, for a discussion).
}
\label{fig:2steps}
\end{figure}

Let us describe the citizen classification here retained to characterize the electoral mobilization of registered voters. For each given election and each specific scale (a municipality, a province, a country, etc.) we distinguish: (1) the total number $N$ of registered Voters; (2) the number $\na$ of Abstentionists, the persons who do not take part to the election; (3) the number $\nv$ of voters, among which (4) $\nn$ Blank and Null Votes (some countries, like Canada and Poland, aggregate Blank and Null votes in an only one term called as Null votes, or Invalid votes, or Spoilt votes) and (5) $\nc$ Votes in favor of candidates or electoral list of choices, also sometimes called Valid Votes (see Fig.~\ref{fig:2steps}). Obviously $\nv = \nc + \nn$ and $N = \na + \nc + \nn$. Note that in Italy, Spain, and Switzerland, electoral data distinguish between Null Votes, $\nnu$, and Blank Votes, $\nwh$. Moreover, only in Spain, \textit{``Votos V\'alidos''} means $\nv - \nnu$, that differs from other countries where ``Valid Votes'' means $\nv - \nnu - \nwh$. In this paper, we consider for all countries that Valid Votes are defined as $\nc = \nv - \nn$. See Section~F in Appendix S1 for more discussion about countries where Blank Votes and Null Votes are distinguished between each other.

As discussed in the following, we characterize the civic involvement of registered voters by the choice between the three possible sates, Abstention, Blank or Null Vote and Valid Vote. The civic involvement of electors is then here measured through the set of the three ratios $\{\pa, \pc, \pn\}$, defined by
\begin{equation}\label{ep} \pa = \frac{\na}{N},\quad \pc = \frac{\nc}{N},\quad \pn = \frac{\nn}{N},\end{equation}
with $\pa + \pc + \pn = 1$. Each election can then be represented by a point in the simplex $\pa + \pc + \pn = 1$, as illustrated on Fig.~\ref{simplex}. Since the number of Blank and Null is typically small, clearly most points lie near the edge $\pn=0$. A second basic observation is that there is a wide dispersion along the axis $\pa - \pc$. Figure~\ref{fig:new} shows the scatter plot of ($\pa,\, \pn$) for French elections since 2000, and the 100 most populated cities. This plot suggests that individual behavior cannot be explained by a sequential binary choice (first to decide to vote or not, and if yes, then to decide to cast a valid vote or not), since this would lead to the absence of correlations between $\pa$ and $\pn$. Hence the {\it electoral involvement} should be viewed through the three possibilities available to the voters: abstention, blank/null votes and votes according to the list of choices. Moreover, Figure~\ref{fig:new} shows that, if there are statistical regularities, they can be seen by considering a convex function of the variables $\pa$, $\pn$. This is what we do below, making use of the entropy function associated to the three quantities $\pa$, $\pc$ and $\pn$.

Previous work~\cite{turnout-stat} has revealed statistical regularities from election to election, and from country to country, when considering the distribution of turnout over municipalities. More precisely, the distribution of the logarithmic turnout rate, $\tau \equiv \ln(\frac{1-\pa}{\pa})$, {\it centered on its mean value}, is remarkably stable over time and across countries for the most populated cities. Similarly, a logarithmic three choices value can be defined, $\tau_3 = \ln\big(\frac{\pc\cdot\pn}{(\pa)^2}\big)$, for which, the same type of regularities can be observed when considering polling stations within municipalities (see Appendix S1, Section D.1). In addition, this analysis of fluctuations confirms the remark in the preceding paragraph, that individual behavior is not well explained by a sequential binary choice (see Appendix S1, Section D.2). However, this analysis of fluctuations does not say anything on the mean values. In this paper, we exhibit another type of regularities, by considering an adequate function ($\tau_3$ has not the appropriate convex properties), and focusing on the values themselves, not only the fluctuations around the means.

\subsection*{The involvement entropy}
We introduce a variable whose value, as we will argue, is appropriate for characterizing the mean civic involvement of the electorate. Viewing the three ratios $\{\pa, \pc, \pn\}$ as probabilities, it is interesting to associate to each election, instead of these three numbers, a single scalar characterizing the probability distribution itself.
One natural quantity associated to a probability distribution is the entropy, $\s$, defined by
\begin{equation}\label{es3} \s (\pa, \pc, \pn) = -\pa \log(\pa)-\pc \log(\pc)-\pn\log(\pn).\end{equation}
Here, and throughout this paper, $\log$ means base-two logarithm ($\log(2)=1$, and the entropy is said to be in units of bits).

Within the framework of Information Theory, where it is called the Shannon entropy, this quantity can be understood as a measure of missing information, or of average surprise, associated to the studied random process~\cite{feldman-cours}. In the context of Statistical Physics, it is the Boltzmann-Gibbs entropy measuring the degree of `disorder' of the system under consideration~\cite{diu}. In the present context, we will refer to $\s$ as the entropy of civic involvement, or ``involvement entropy'', and consider it as a measure of disorder vs. order in the civic involvement at a collective level. Indeed, it is a `macroscopic' or collective measure about the civic involvement of an electorate, and not the measure of the civic mobilization of individual citizen -- i.e., we do not claim that it corresponds to the behavior of a representative citizen. It can be measured at any scale of aggregate data, e.g. for a municipality, a province, or a whole country. For instance, the involvement entropy of a municipality, $\s$, is given by Eq.~(\ref{es3}) where the three ratios $\{\pa, \pc, \pn\}$ are the ratios of, respectively, the number abstentionists, $\na$, valid votes, $\nc$, and blank and null votes $\nn$, over the total number $N$ of registered voters in the considered municipality.  

Let us explain more what we mean by `order/disorder', and how this is reflected by the entropy value. We consider that a civic involvement shows an `ordered' state if one of the three ratios is very close to one (hence the two others very small). A `disordered' state corresponds to having all three ratios of similar values. Within this viewpoint, no particular role or importance is assigned to any one of the three possible cases, abstention, blank/null, valid vote. The involvement entropy $\s$, a positive or null quantity, provides a well defined way to quantify the degree of disorder: the larger the entropy, the larger the disorder. The maximum order is obtained when one of the ratios is equal to unity (and then the two others are equal to zero), in which case $\s=0$. In contrast, the maximum disorder corresponds to an equipartition of these 3 ratios, that is $\pc=\pa=\pn=1/3$, in which case the entropy takes its maximal possible value, $\s=\log(3)\simeq1.58$.

As an illustration, consider the elections for the Mayor in the French municipalities.
It is well known (at least in France) that participation to elections in small municipalities is typically larger than in large cities, for social reasons -- for instance, in small municipalities where everyone knows every one else, not going to the polling station will become common knowledge. Such social enforcement of the civic involvement might be at the root of an increase of the number of abstentionists with population size: the ratio $\pa$ of abstentionists is typically very low for small municipalities, and increases with the municipality size, $N$. One then expects an increase of the involvement entropy with municipality-size: this is indeed what we observe for the elections for 
the 2001 and 2008 first round (elections for which we have the data for all the municipalities), as illustrated on Fig.~\ref{fmayor}. We can say that the electorate is very ``ordered" (in terms of its civic involvement) for low municipality-size, and gets more ``disordered" with increasing $N$. This involvement entropy increase is observed until a threshold population size value, at which the electoral rule changes: the citizen has a larger number of possible voting choices in municipalities with a number of inhabitants smaller than $3500$, than in more populated municipalities. (It is allowed for citizens living in municipalities with less than $3500$ inhabitants, to combine candidates from different opposite lists, or to add new names from citizens who are not officially candidates.) Remarkably, above this critical size, the involvement entropy becomes essentially independent of the population size: one has a plateau, at $\s$ slightly above $1$, despite variations in $\pa$, $\pc$ and $\pn$. As we will see throughout this paper, this particular value of involvement entropy, $\s \approx 1$, shows up as a typical value in modern elections for most populated cities. 

\begin{figure}[h!]
\centering
  \includegraphics[width=0.45\linewidth, height=6cm, clip=true]{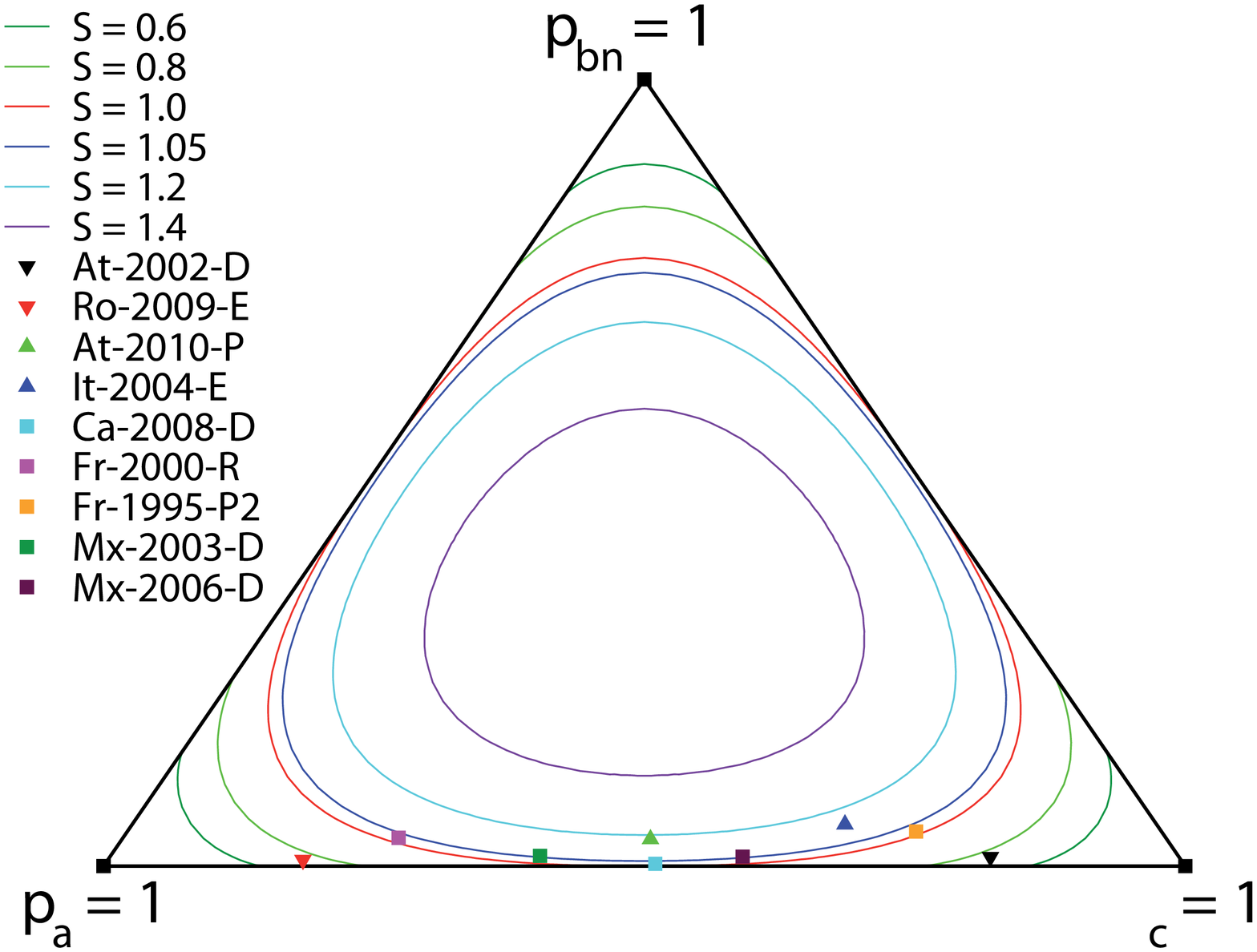}
  \caption{{\bf Simplex $\pa + \pc + \pn = 1$}, in which any given election, for the most populated municipalities, can be represented by a point, as illustrated by the symbols corresponding to particular elections of our data set (see Appendix S1, Section~A, for details). The continuous curves are lines of constant involvement entropy value, drawn for values ranging from $\s=0.6$ to $\s=1.4$. See text for At-2002-D, Ro-2009-E, At-2010-P and It-2004-E. For Ca-2008-D: $\pa\simeq0.49$, $\pc\simeq0.51$, $\pn\simeq0.003$ and $\s\simeq1.02$; for Fr-2000-R: $\pa\simeq0.71$, $\pc\simeq0.25$, $\pn\simeq0.036$ and $\s\simeq1.02$; for Fr-1995-P2: $\pa\simeq0.23$, $\pc\simeq0.73$, $\pn\simeq0.044$ and $\s\simeq1.01$; for Mx-2003-D: $\pa\simeq0.59$, $\pc\simeq0.40$, $\pn\simeq0.013$ and $\s\simeq1.04$; and for Mx-2006-D: $\pa\simeq0.40$, $\pc\simeq0.58$, $\pn\simeq0.012$ and $\s\simeq1.04$.}
\label{simplex}
\end{figure}

\begin{figure}[h!]
\centering
  \includegraphics[width=0.45\linewidth, height=5cm, clip=true]{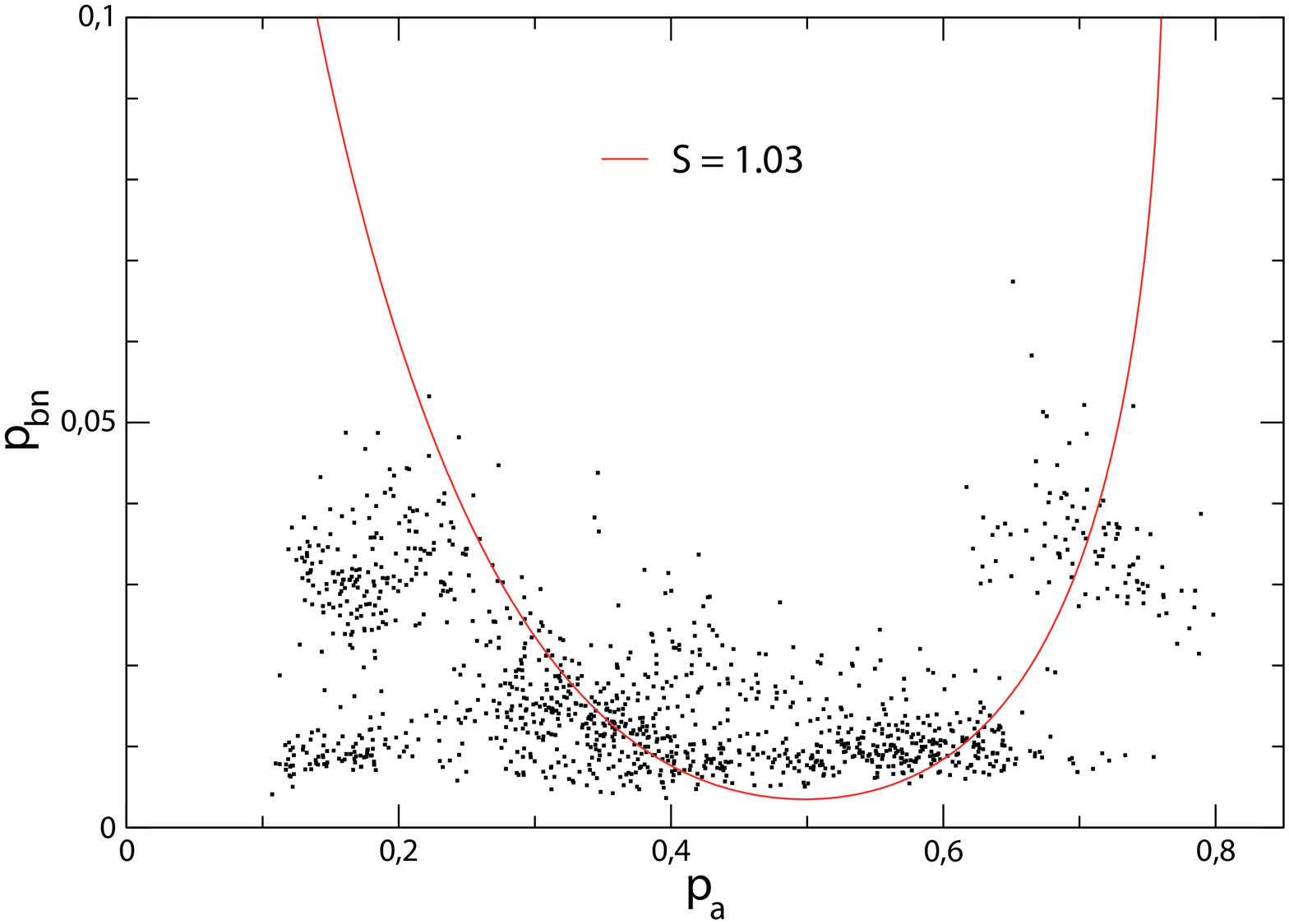}
  \caption{{\bf Scatter plot of ($\pa,\, \pn)$ of the 100 most populated French cities over elections since 2000.} The convex function $\s$ (here $\s=1.03$) is also plotted as a guide view.}
\label{fig:new}
\end{figure}

Let us give other illustrations. A great order of the electorate is provided by: (1) the population of registered voters is highly polarized: there is an important difference between $\pa$ and $\pc$ ($\pa\ll\pc$ or $\pa\gg\pc$); and (2) blank and/or null votes are very few, that is $\pn$ is very small. Such cases of small entropies are, e.g., the 2002 Austrian Chamber of Deputies election for which $\pa\simeq0.17$, $\pc\simeq0.81$, $\pn\simeq0.011$ and $\s\simeq0.73$; the 2009 European Parliament election in Romania, with $\pa\simeq0.81$, $\pc\simeq0.18$, $\pn\simeq0.008$ and $\s\simeq0.73$. Conversely, a great disorder of the electorate results from: (1) the population of registered voters is not very polarized, that is $\pa$ and $\pc$ are not very different; and (2) blank and/or null votes are relatively important, that is $\pn$ is not too small. For instance, the 2010 Austrian Presidential election has $\pa\simeq0.48$, $\pc\simeq0.49$, $\pn\simeq0.034$ and $\s\simeq1.16$; and the 2006 European Parliament election in Italy has $\pa\simeq0.29$, $\pc\simeq0.66$, $\pn\simeq0.053$ and $\s\simeq1.11$. Note that these values come from great town values (see the SI, Tab.~S1), whereas $\s$ is more spread out in small municipalities (see Fig.~\ref{fmun}). Finally, one finds that the involvement entropy $\s$ has a value frequently very near $1.0$. For example, the 2008 Canadian Chamber of deputies election, the 2000 French referendum, the 1995 French second round Presidential, and the 2003 and 2006 Mexican Chamber of deputies elections (see Fig.~\ref{simplex} and Tab.~S1 in the SI). In all these examples, despite an important diversity in $\pa$ values, $\s$ lies within $1.01$ and $1.04$, showing that the electorate polarization is somewhat halfway between order and disorder. Note that $\s=1$ is the entropy associated to the tossing of a fair coin. In the present context, it would be exactly obtained for elections with $\pa=\pc=50\%$ and $\pn=0$.

\section*{Results}

\subsection*{Stylized fact: The common occurrence of $\mathbf{\s\approx 1}$} 

We have computed the involvement entropy $\s$ for all the elections of our data set, at different scales. First we find that, most often, it depends on the municipality-size $N$. To analyze this size dependency, we spread out municipalities data over samples with respect to the municipality population-size. In each sample, municipalities have roughly the same number of registered voters. The number of municipalities per sample is of order $100$, except for France in which case this number is $200$ (because France has much more municipalities than the other countries studied in this paper). We denote by $\smoy$ the average over all municipalities inside a sample of the involvement entropy $\s$. This average $\smoy$ is plotted in Fig.~\ref{fmun} as a function of the number of registered voters, $N$.  

In this paper, $\overline{X}$ means the average value of the considered value, $X$, over all municipalities, around 100, or 200 for France, in a given sample where municipalities have roughly the same number of registered voters, $N$; e.g. $\smoy$, $\overline{\pa}$, $\overline{\pn}$, etc. Average values and standard-deviations do not take into account extreme values in order to remove some electoral errors, etc. Electoral values greater than \textit{5 sigma} are not taken into account. For instance let 100 municipalities of size $\approx N$ (as in Fig.~\ref{fmun}), each one has a civic involvement entropy $\s_i$ ($i=1,2,...,100$). First, $\langle\s\rangle$ and $\sigma$ are the average value and the standard-deviation of $\s$ over these 100 municipalities. Next, the final average value $\smoy$ and the final standard-deviation over this sample of 100 municipalities are only evaluated for municipalities, $i$, such that $|\s_i - \langle\s\rangle| < 5\;\sigma$.

Let us give the 1995 French second round Presidential election (Fr-1995-P2) as an example. A relatively ordered civic electorate involvement is observed for the smallest population-size municipalities, with $\smoy\simeq0.7$. The mean involvement entropy then increases with municipality size, for sizes up to $N\sim 10000$. For the most populated municipalities, that is above this threshold value in population-size, a saturation occurs: the (average) civic disorder of the electorate becomes independent of municipality-size, with $\smoy\approx 1$.

Next we consider the time evolution of the involvement entropy at a large scale (country, province, {\it canton}, etc.). When the scale of aggregate data is lower than the national one, each value of the involvement entropy for one election is equal to a weighted (by population-size) mean value of involvement entropies at lower scale (province, {\it canton}, etc.). (See Appendix S1, Section~A, and Tab.~S2 in the SI for more details.) Fig.~S1 in the SI plots the involvement entropy of each election at large scale, for each country over all elections (according to its nature) as a function of time, and Fig.~S2 in the SI shows how $\pa$ and $\pn$ evolve in time for Chamber of Deputies election in each country. Nevertheless a rapid evolution in time of $\s$ can be seen in a different way. First, for each country, elections are ordered according to their year; half of them, the more ancient ones, are gathered into one group, and the other half, the more recent ones, are gathered into another group. Next, we aggregate over countries, the ``old'' elections on one side, the ``recent'' ones on the other, getting a total of 321 elections split into two groups with roughly the same number of elections in each one. Although recency is here country specific, the aggregated group of recent elections corresponds more or less to those occurred since the 70s. The histograms of the involvement entropy $\s$ are compared for these two groups on Fig.~\ref{fhisto-dyn}. The histogram for the group of the more recent elections shows a sharp peak at $\s\approx 1$, whereas the group of the older elections has a broad distribution. This temporal evolution occurs in parallel with a significant decrease of ``highly ordered'' elections (in the civic involvement point of view). In other words, nowadays there are few elections with a small civic involvement entropy, $\s$ (say e.g. $\s\lesssim 0.8$), but there are a lot of elections with $\s\approx 1$.

Finally, Fig.~\ref{fE} shows, for all the European Parliament Elections, how the involvement entropy of municipalities depends on population-size (like in Fig.~\ref{fmun}), and the time evolution at the national or provincial scale (like Fig.~S1 in the SI).

\subsection*{What the common occurrence is}

As already said, Fig.~\ref{fmun} shows the remarkable fact that, for each studied country, in modern elections the involvement entropy of highly populated municipalities is very frequently roughly equal to $1$. This common value, $\s\approx 1$, for high population-size municipalities is particularly striking when one looks at European Parliament Elections (see Fig.~\ref{fE}-a). See also Table~\ref{tdern} for a rapid overview and basic statistics per country about involvement entropies and population size of the $\approx 100$ most populated municipalities. There are however noticeable exceptions, notably the Italian case on which we will come back later (Section Discussion). In any case, we have now to better specify what we mean by $\s\approx 1$ and show more quantitatively in which way it is a common property of modern elections. This is done by gathering data over all elections after 2000 (after 2000 in order to take into account evolution in time of the involvement entropy as stressed by Fig.~\ref{fhisto-dyn} and Tab.~\ref{tdern}). Fig.~\ref{fhisto-size}-d plots the resulting histograms of the involvement entropy restricted to $\approx 100$ most populated municipalities, for different countries or ensemble of countries. Moreover, Fig.~\ref{fpic} shows respectively the minimal length interval of $\s$, $\pa$, $\pc$ and $\pn$ which contain $50\%$ of events (those plotted in Fig.~\ref{fhisto-size}-d). These two figures show a common sharp peak at a value of $\s$ close to $1$. The involvement entropy appears to be mainly in the range $0.98\lesssim\s\lesssim1.08$, which can be taken as the definition of $\s\approx1$ in this paper. Note that this definition is applied to the most populated municipalities. At large scale, the involvement entropy depends on the the way that data are aggregated (at national, province, etc. scale), and it is a little bit greater than $\smoy$ for the most populated municipalities. Nevertheless the involvement entropy measure at large scale approximately reflects how the most populated municipalities do, because an important ratio of population live in the $\approx 100$ most populated municipalities (as seen in Tab.~\ref{tdern}).

It is important to stress that the common occurrence $\s\approx 1$ appears (1) as a property of high populated municipalities, (2) and also in a recent time. See Fig.~\ref{fhisto-dyn}, or Fig.~S1 in the SI, as an indication of the latter point. For the first point, Fig.~\ref{fhisto-size} shows the histograms of the entropy for different municipality sizes. Compared with histograms of the most populated municipalities (Fig.~\ref{fhisto-size}-d), histograms of lower municipality-size appear: (1) much less peaked (apart from Polish elections), and (2) not peaked at the same common-value. Moreover, it is only for the larger sizes that all the histograms become very similar, suggesting the convergence to a universal histogram at large sizes. Let us bear in mind (cf. Tab.~\ref{tdern}) that the sample of the $\approx 100$ most populated municipalities in Austria is, on average, much less populated than the ones of the four other countries or ensemble or countries. (Taking into account the $\approx 50$ Austrian municipalities per sample provides, for the most populated sample, an histogram of $\s$ much centered on $\s\approx 1$ than the one of $\approx 100$ municipalities (see Fig.~S3 in the SI).) In other words, the Austrian sample of the $\approx 100$ most populated municipalities is not so comparable to the four other ones, and does not accurately reflect a typical behavior in large populated municipalities (especially since the civic involvement can significantly depend on the population size as it is shown in Fig.~\ref{fmun}). Lastly, the choice of the number (here $100$) of most populated municipalities is only for statistical convenience and does not affect the results (see e.g. Fig.~S3 in the SI which is similar to the Fig.~\ref{fhisto-size}-d, but for the sample of $50$ or $200$ most populated municipalities).

Now, let us better quantify this sharp and common peak for the most populated municipalities. First, Fig.~\ref{foverlap}-a plots the smallest distance $(\s_{sup}-\s_{min})$, such that $50\%$ of events are included into the set $[\s_{inf},\,\s_{sup}]$, with respect to the relative municipality size. This confirms that (apart from Polish elections) distributions of $\s$ get more peaked when the population size increase, and specifically for the most populated municipalities. (The same features also appear by considering minimal distances which contain $25\%$ or $10\%$ of events. This is in agreement of the robustness of this trivial method.) Moreover (apart from the Austrian elections) the minimal distance $(\s_{sup}-\s_{min})$ appears to converge to a common value, this only for the most populated sample (see also Fig.~\ref{fpic} for $\s_{inf}$ and $\s_{sup}$ for this latter sample). Next, in order to quantify the common peak phenomenon, we calculate the overlap between distributions of $\s$ for municipalities as a function of the relative population size (see Fig.~\ref{foverlap}-b). The overlap between $n$ distributions of $\s$, with probability density functions (pdf) $f_i(\s)$, $i=1,2,\cdots,n$, is defined as $\mathcal{O}_n = \int \min\big[f_1(\s),\,f_2(\s),\,\cdots,\,f_n(\s)\big]\, \mathrm{d}\s \,.$ Fig.~\ref{foverlap}-b shows an increasing overlap between distributions when the population size increases, and specifically for the most populated municipalities. This confirms that the distributions of $\s$ get more and more similar as the relative municipality-size increases, with (sharp) peaks becoming identical for the most populated municipalities.

\subsection*{What the common occurrence is not}

We claim that this common most frequent value, $\s\approx1$ for the most populated municipalities, is not a mere statistical artefact. More precisely, we claim that:
\begin{itemize}
\item[(1)] it is not a direct consequence of the law of large numbers, which, for data aggregated at the scale of large municipalities, would give a systematic result;
\item[(2)] it is not a result of `pure chance', that is a bias in the data due to random events, or an accidental bias in the collected data;
\item[(3)] it does not only result from having $\pa$ and $\pc$ neither around $50\%$ nor around a common value: there is a wide range of $\pa$ values for which $\s\approx1$ is observed;
\item[(4)] it does not result from having a small proportion of Blank and Null Votes.
\end{itemize}
In support of the two first points, we show below that there are robust properties which cannot be explained by the pure chance or the large number hypotheses. About the two last points concerning the ranges of $\pa$ and $\pn$ values, we show that, even if the distributions of $\s$ could be peaked for a relatively broad distribution of $\pa$ and small values of $\pn$, this, (1), cannot alone explain why the distributions of $\s$ for the most populated towns are so much narrowed and, (2), in addition, have their peak at a common value of $\s$. The next three sections detail these claims.

\subsubsection*{Against a randomness or large number artefact}
We note three facts that goes against a pure chance or large number hypotheses.\\

(i) $\s\approx 1$ is specific to modern elections. Indeed (apart from Swiss {\it Votations} discussed in Section Discussion) this common value $\s\approx 1$ appears recently, and at different times for different countries -- and different elections --: in the 70's or 80's in France, 80's in Germany, 90's in Canada, 2000's in Czech Republic, etc (cf. Fig.~S1 in the SI). Moreover, there is no systematic way in which recent convergence to $\s\approx 1$ appears in time. $\s\approx 1$ may be reached as well from inferior values (e.g. Chamber of Deputies elections in Canada, Czech Republic, etc., in Fig.~\ref{fE}-b) than from superior values (e.g. European Parliament in France in Fig.~\ref{fE}-b). Lastly, in a given country, some kind of elections provide at large scale $\s\approx 1$ since their coming (e.g. European Parliament elections ), and for some other ones, $\s\approx 1$ seems (actually) to be an attractor point in time (see e.g. Chamber of Deputies elections in Canada, Czech Republic, France, Switzerland, etc. in the SI, Fig.~S1).   

\begin{figure}[h!]
  \centering
\includegraphics[width=0.45\linewidth, height=6cm, clip=true]{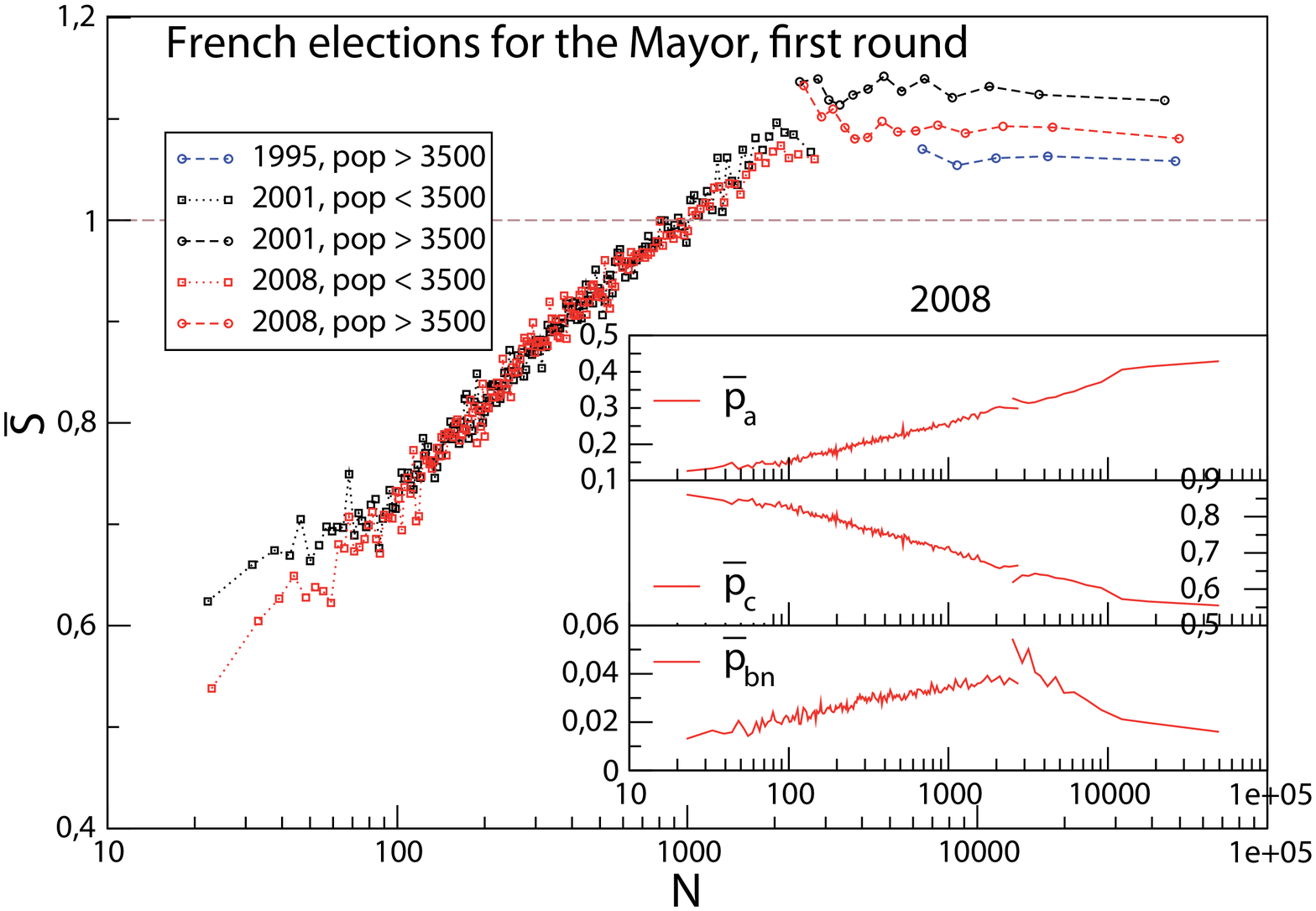}
\caption{{\bf Average values $\smoy$ of the involvement entropy of municipalities, $\s$, as a function of the number of registered voters $N$, for the first round of Mayor elections in France.} There are two kinds of voting rules, which depend on the population-size more or less than 3500 inhabitants (see text). Inset shows average values of $\pa$, $\pc$ and $\pn$ as a function of $N$ for the 2008 \textit{municipales} elections (which lead for high population municipalities to a plateau of $\s$ despite variations in $\pa$, $\pc$ and $\pn$). For each $N$, average values, $\smoy, \overline{\pa},\overline{\pc}, \overline{\pn}$, are evaluated over $\approx 200$ municipalities of size $\approx N$.}
\label{fmayor}
\end{figure}

\begin{figure*}[h!]
\centering 
\includegraphics[width=16cm, height=10.5cm, clip=true]{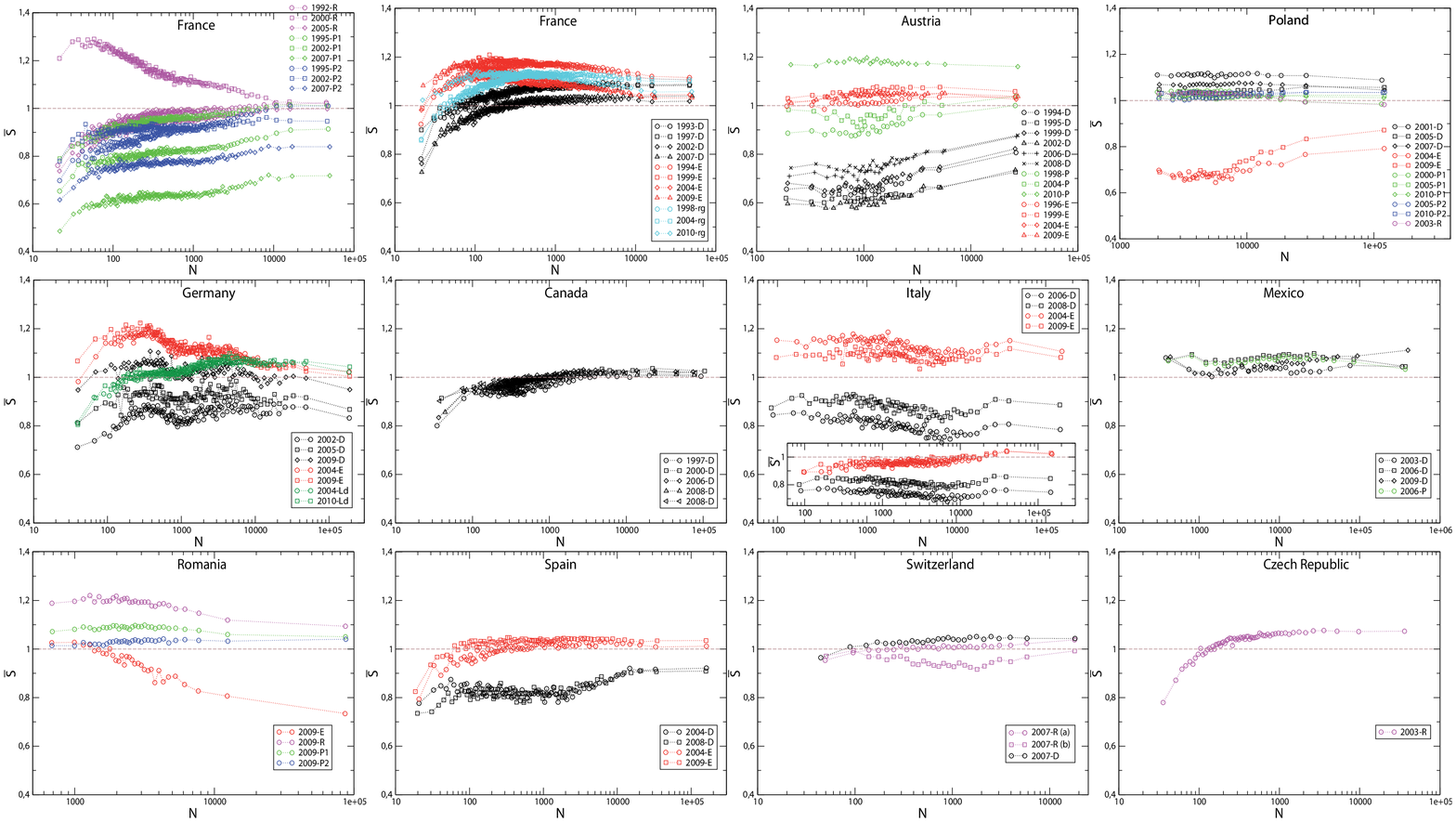}
\caption{{\bf Mean values $\smoy$ of the involvement entropy for municipalities, as a function of the number of registered voters $N$.} Each point results from an average over a sample of $\approx$ 100 (200 for France) municipalities of size $\approx N$. Italian graph inset shows a variant of $\s$ where Blank Votes are grouped with Valid Votes (see Appendix S1, Section~F, for a deeper discussion). See the Appendix S1, Section~A, for more details on the data.}
\label{fmun}
\end{figure*}

\begin{figure}[h!]
  \centering
\includegraphics[width=0.45\linewidth, height=6cm, clip=true]{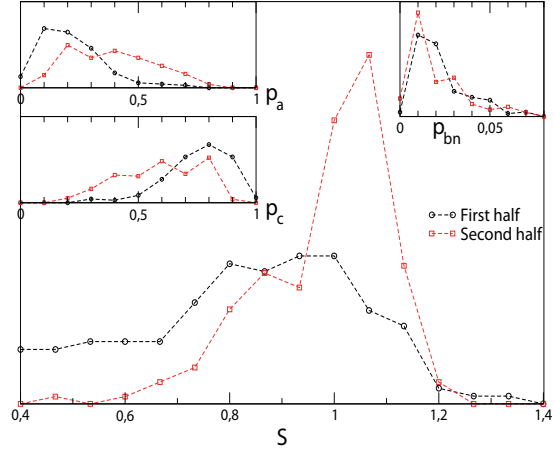}
\caption{{\bf Evolution in time of involvement entropy, $\s$, at large scale} (national, provincial, etc.) of 321 elections (see Appendix S1, Section~A, for more details), apart from Swiss referendums. For each country, electoral results are equally divided into two groups: those which occurred at the first period in time and at the second one. Histograms of $\s$ (and $\pa$, $\pc$ and $\pn$ in the insets) show the involvement entropy of the first and second group over all countries. Fig.~S1 in the SI plots for each country the whole of elections, and also Fig.~S4 in the SI for scatter plots $(\pa,\,\pn)$ of these elections, but at national aggregate scale.}
\label{fhisto-dyn}
\end{figure}

\begin{figure*}[h!]
\centering
\includegraphics[width=16cm, height=6cm, clip=true]{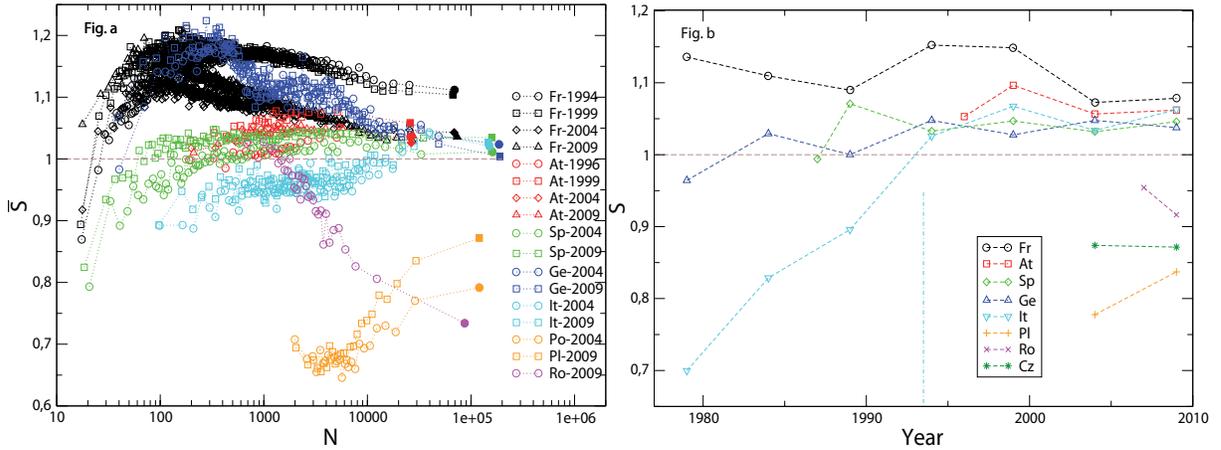}
\caption{{\bf Mean involvement entropy for European Parliament elections.} Fig.~\ref{fE}-a (left panel): same elections as those shown in Fig.~\ref{fmun}; here for all countries, including France, averages are over $\sim 100$ municipalities. Fig.~\ref{fE}-b (right panel): same elections as those shown in Fig.~S1 in the SI;
the vertical dashed line indicates the year of the abolishment of compulsory voting in Italy. Here, Italian Blank Votes, $\nwh$, (but not Null Votes, $\nnu$) are grouped with votes in favor of lists of candidates (see Appendix S1, Section~F, for more discussion).}
 \label{fE}
\end{figure*}

\begin{figure*}[h!]
\centering
\includegraphics[width=16cm, height=7cm, clip=true]{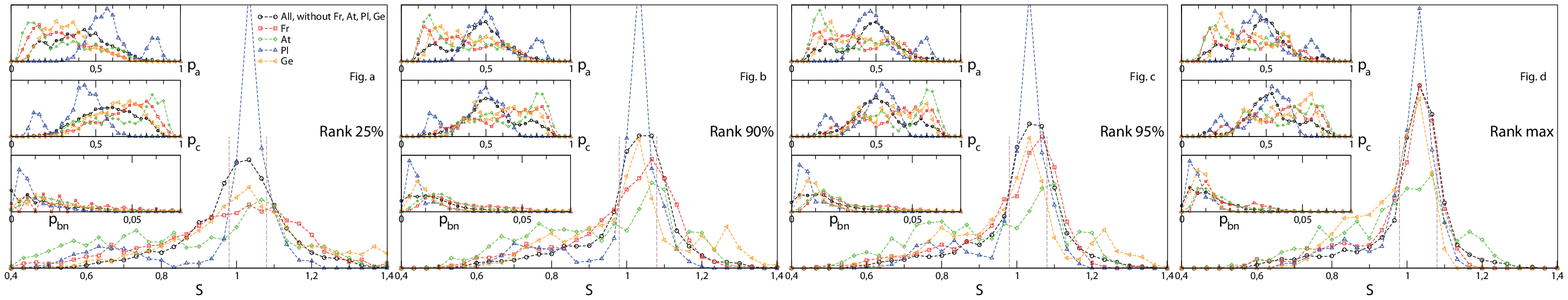}
\caption{{\bf Histograms of involvement entropy, $\s$, with respect to the relative municipality-size bin} over all analyzed since 2000. There are 12 French elections, 7 Austrian elections, 11 Polish elections, 7 German elections and 24 for others countries (included in one curve, with no more than 4 elections per country). Municipalities of each country are divided into bins (of $\approx 100$ municipalities) with respect to their municipality-size (see e.g. Fig.~\ref{fmun}). For instance, `Rank $25\%$' (Fig.~\ref{fhisto-size}-a) means the bin whose population-size rank is the twenty-fifth per cent with regard to the sample of the most populated municipalities (Fig.~\ref{fhisto-size}-d) of the considered country. Insets: histograms of corresponding $\pa$, $\pc$ and $\pn$. $\s=0.98$ and $\s=1.08$ are plotted in dashed lines and all the scales axis are similar from one plot to another one.}
\label{fhisto-size}
\end{figure*}

\begin{figure*}[h!]
\centering
\includegraphics[width=16cm, height=4cm, clip=true]{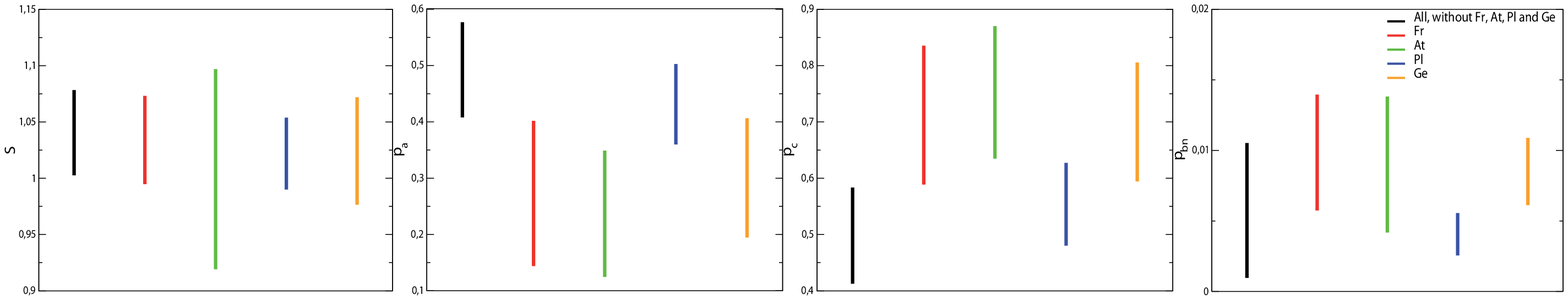}
\caption{{\bf Minimal intervals containing $50\%$ of events}, of the involvement entropy $\s$ and ratios $\pa$, $\pc$ and $\pn$ of the 100 most populated municipalities over all elections since 2000. See Fig.~\ref{fhisto-size}-d for the related histograms.}
\label{fpic}
\end{figure*}

\begin{figure*}[h!]
\centering
\includegraphics[width=16cm, height=6cm, clip=true]{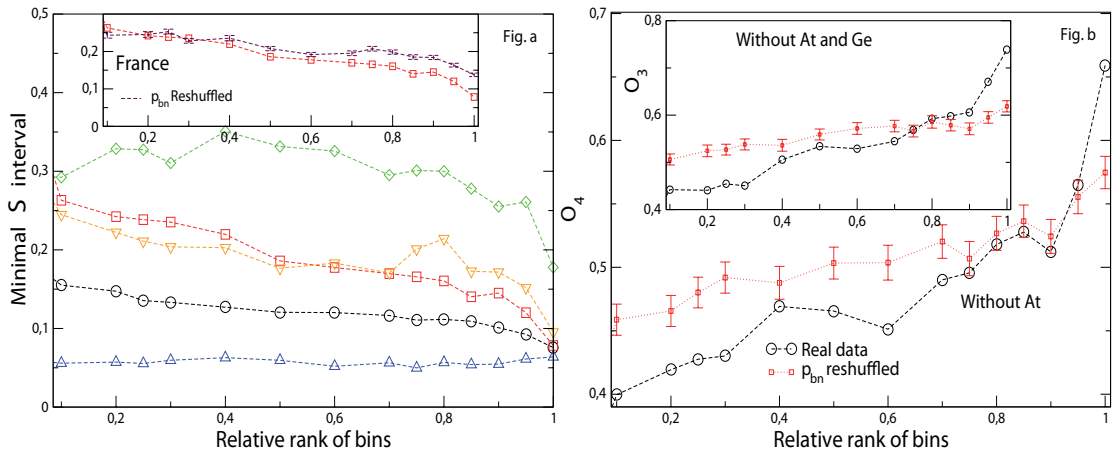}
\caption{{\bf Quantitative evidence of the sharp and common peak of $\s$ for the most populated municipalities.} Considered elections, the way that bins are ranked, countries or groups of countries and legends are the same as in Fig.~\ref{fhisto-size}. Left (\ref{foverlap}-a): Minimal interval $(\s_{sup}-\s_{inf})$, which encapsulates $50\%$ of events, with respect to the relative population size. Right (\ref{foverlap}-b): Overlap $\mathcal{O}_4$ between 4 distributions of $\s$ of municipalities as a function of their relative municipality-size (see text for the definition of $\mathcal{O}_n$). The inset shows in the same manner overlap $\mathcal{O}_3$ between 3 distributions of $\s$ ((1): all without At, Fr, Ge and Pl; (2): Fr; (3): Pl). Some curves obtained from reshuffling $\pn$ of municipalities (inside one country or ensemble of countries), while $\pa$ is not modified, are also plotted.}
\label{foverlap}
\end{figure*}

(ii) $\s\approx 1$ is only observed for large populations (and there is no common-value for smaller municipality sizes) as it is shown in Fig.~\ref{foverlap}; and there is sometimes  a plateau with a lower value of $N$ which both depend on the election and on the country (e.g. $\approx$ 3000 in Canada and Czech Republic, 10000 in France for referendums, etc., in Fig.~\ref{fmun}). Moreover, there is no systematic way in which convergence to $\s\approx 1$ occurs as the population size increases. $\s\approx 1$ may be reached as well from inferior values (e.g. Fr-1995-P2, Sp-2004-E and Sp-2009-E) than from superior values (e.g. Fr-2000-R, Ge-2004-E and Ge-2009-E in Fig.~\ref{fmun}). Lastly, $\s\approx 1$ may be reached from a discontinuous transition when voting rule (which depends on the population size of municipalities) changes. This occurs for the two French local elections for the Mayor (see Fig.~\ref{fmayor}), which are the only one elections of our database where there is this electoral rule change.

(iii) The shape of distributions of $\s$ for large municipality sizes does not result from a statistical bias due to large numbers: creating artificial high populated municipalities, by means of aggregating large amount of citizen choices who live in small and different municipalities, does clearly not yield a distribution peaked near $\s\approx 1$ (see Appendix S1, Section~C, and Fig.~S6 in the SI for more details). 

Finite-size-effects, that is the effect of aggregating data at different scales, are considered more thoroughly in Appendix S1, Section~C, comparing ballot box scale with municipality scale. This section also discusses more the issue of statistical effects that could be due to large numbers.


\subsubsection*{Ranges of variation of $\pa$ and $\pn$}

On one side, while $\pn$ does not radically change in time at large scales, $\pa$ has increased during last decades in most countries (see e.g. insets of Fig.~\ref{fhisto-dyn} and Fig.~S2 in the SI). On the other side, $\pn$ is known to decrease when the population-size of municipalities $N$ increases, as it was discussed in the Section Introduction. Let us thus first consider the possibility that the common occurrence $\s\approx 1$ could be a consequence of these two facts: $\pa$ is not too small (for example, if $\pa\lesssim 0.227$, then it is no more possible to get $\s=1$) and, {\it independently}, $\pn$ is small.

\begin{figure*}[h!]
\centering
\includegraphics[width=16cm, height=6cm, clip=true]{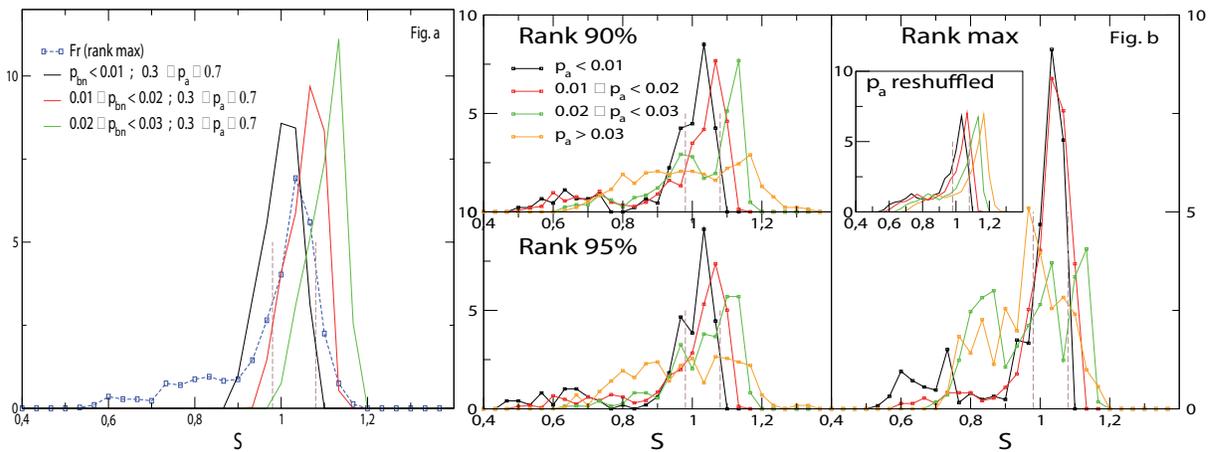}
\caption{{\bf Relative importance of $\pn$ on the distribution of $\s$.} Left (\ref{fmultipbn}-a): Distribution of $\s$ from flat distributions of $\pa$ (where $\pa\in[0.3 ;0.7]$) and $\pn$. The pdf of $\s$ can be peaked for a relatively broad distribution of $\pa$ and a small range of $\pn$, but the peak of the distribution of $\s$, which depends on $\pn$ values, is not necessarily centered on $\s\approx1$. Histogram of $\s$ over the $\approx 100$ French most populated municipalities (the same as in Fig.~\ref{fhisto-size}-d) is also given as a guiding view. Right (\ref{fmultipbn}-b): pdf of $\s$, for different ranges of $\pn$, over $\approx 100$ French municipalities, which depend on their relative population-size (like in Fig.~\ref{fhisto-size}-b, c, d). Histograms of $\s(\pa,\pn)$ from reshuffled $\pa$, while $\pn$ remain unchanged, are also given.}
\label{fmultipbn}
\end{figure*}

We give three arguments against this assertion.
(i) First, we plot on Fig.~\ref{fmultipbn}-a histograms of $\s$ resulting from a flat and broad distribution of $\pa$, and a flat distributions of $\pn$ (with small values). Each histogram corresponds to a different choice of the range of (small) $\pn$ values. To better understand this point, let $S_2(\pa) = -\pa \log(\pa) -(1-\pa) \log(1-\pa)$ which has a maximal value, $S_2=1$, for $\pa=0.5$. Moreover, when $\pn = 0$, $S_2$ is equal to the involvement entropy, $\s$ defined in Eq.~(2), i.e. $S_2(\pa)=\s(\pa,\pn=0)$. Hence, relatively small variations of $\pa$ around $0.5$ and very small values of $\pn$ lead to $\s\simeq 1$.) The result is indeed a set of peaked histograms. However, these distributions of $\s$ are neither necessarily centered on $\s\approx1$ nor centered at a common peak. 
 
(ii) Second, we emphasis the specificity of most populated municipalities.
Fig.~\ref{fmultipbn}-b plots for French data (where the tested phenomenon is clearer) distributions of $\s$ selecting elections for which $\pn$ belongs to specific ranges of values. Moreover theses distributions are also plotted according to the population-size of municipalities. It is only for the most populated municipalities that the distributions of $\s$ for different ranges of $\pn$ are roughly peaked at the same value $\s\approx 1$ (with a very good agreement for $\pn \in[0,\,0.01[$ and $\pn \in[0.01,\,0.02[$). Moreover, for a lower population-size, e.g. with a relative rank of $90\%$, it is interesting to note that distributions of $\s$ for different ranges of $\pn$ (apart from the $\pn\geq 0.03$) share the same features as in Fig.~\ref{fmultipbn}-a, i.e. distributions are peaked in different values. (To have a more detailed view, Fig.~S5 in the SI shows scatter-plots $(\pa,\,\pn)$ for the municipalities taken into account in Fig.~\ref{fmultipbn}-b.).

(iii) Third, there is actually a wide disparities in the ranges of $\pa$ and $\pn$ between different countries or group of countries. One can see in Fig.~\ref{fpic} how, (1), France and, (2), all countries without At, Fr, Ge and Pl, can reach the common $\s$ peak, despite largely different ranges of $\pa$, $\pc$ and $\pn$. In other words, 
the ranges of ratios $\pa$, $\pc$ and $\pn$ are not sufficiently similar between countries or ensemble of countries to explain why the distributions of $\s$ for the most populated municipalities share a sharp peak at a common value of $\s$.

\subsubsection*{Implied correlations between $\pa$ and $\pn$}

Hence, it seems difficult to explain the common value $\s\approx 1$ for the most populated towns as a consequence of having {\it independently} $\pn$ small and $\pa$ in a given particular range. The observation of a common peak around $\s\approx 1$ thus implies the existence of specific correlations between $\pa$ and $\pn$. 

To test this conclusion, we consider surrogate data obtained by reshuffling the ratios $\pn$ from one municipality (or country) to another one, while $\pa$ is kept unchanged (and then $\pc$ is deduced from $\pc=1-\pa-\pn$). Note that the marginal distributions of $\pa$ and $\pn$ remain unchanged by this reshuffling procedure, whereas their correlations are destroyed. We use this method twice: first, (i) contrasting recent and old elections, and second, (ii), considering the dependency in municipality size.
In the following, reshuffling results are shown as average values over 1000 realizations, and the corresponding standard-deviations are plotted as error bars.

\begin{figure}[h!]
\centering
\includegraphics[width=0.45\linewidth, height=6cm, clip=true]{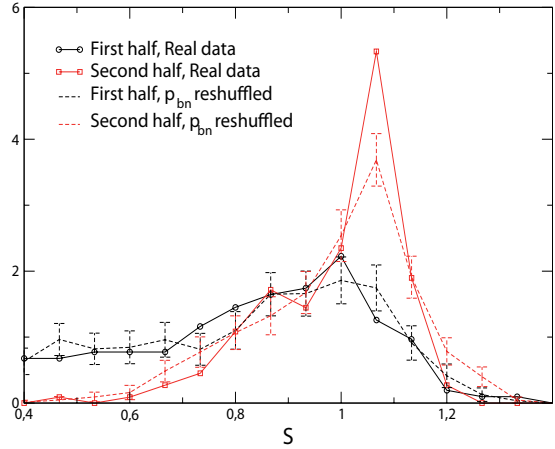}
\caption{{\bf Relative importance of reshuffling $\pn$ on $\s$}. Analyzed elections and the manner that elections are divided into two groups are the same than in Fig.~\ref{fhisto-dyn}. Nevertheless, here, each election is aggregated at national scale, i.e. $\s$ is directly evaluated from the set $\{\pa, \pc, \pn\}$ at the national scale. In these figures, surrogate $\s(\pa,\pn)$ data, consist in reshuffling $\pn$ from one election to another one in the same group, while $\pa$ is not modified. Surrogate curves result from the average of 1000 realizations, and standard-deviations are plotted as error bars.}
\label{fhisto-dyn-mel}
\end{figure}

(i) Figure~\ref{fhisto-dyn-mel} shows, at national scale and for two periods of time, how the distributions of $\s$ change under this reshuffling. $\pn$ are reshuffled within the same group of elections. For the first period in time, the real distribution of $\s$, which is not peaked near $\s\approx 1$, and the surrogate one are not very different between themselves. By contrast, the distributions are notably different for the second period. Moreover, the main difference concerns the peak near $\s\approx 1$. The peak of the surrogate data distribution is less sharp than the one of the real data. This is particularly interesting since $\pn$ is roughly distributed in the same manner between the two relative periods in time (see insets of Fig.~\ref{fhisto-dyn} or scatter-plots $(\pa,\,\pn)$ of Fig.~S4 in the SI). The widening of the surrogate distribution of involvement entropy near the peak $\s\approx 1$ can be seen as a sign that there are correlations between $\pa$ and $\pn$ which enforces the occurrence of $\s\approx 1$. 

(ii) From a qualitative point of view, the reshuffled data have a peak of $\s$ values which is less narrow than for the real ones, a discrepancy which increases with municipality size, as can be seen for the French data on the inset of Fig.~\ref{foverlap}-a, and on the scatter-plots $(\pa,\,\pn)$ on Fig.~S5 in the SI. In addition, the distributions of $\s$ obtained for the reshuffled data are not as well peaked at a common value as it is the case for the real data ones. Quantitatively, for the French data, the Kolmogorov-Smirnov distance between the distributions of real and reshuffled data is significantly larger for the most populated municipalities, with a distance that allows one to reject the hypothesis that the two distributions are similar (indeed the Kolmogorov-Smirnov distance is then $3.0\pm0.2$, while 1.6 corresponds to $\approx 1\%$ probability that the two distributions coincide). Moreover, Fig.~\ref{foverlap}-b shows that overlaps between different distributions of $\s$ resulting from reshuffled $\pn$ is smaller than for real data, and this only when municipality-sizes are high, or even only for the most populated municipality sample: the reshuffling suppresses the high increase of overlaps which is observed on real data for the sample of the most populated municipalities. 

We can thus conclude that there is a specific property for the most populated municipalities, which is not encapsulated by considering $\pa$ and $\pn$ as independent variables.

\section*{Discussion}
We suggest that the common value $\s\approx 1$ of the entropy, which appears recently in high populated municipalities, reveals an emerging collective behavioral norm characteristic of citizen involvement in modern democracies, and we propose to call it a `weak law' on recent electoral behavior among urban voters. Signs of existence of this possible norm can not only be seen notably by the greatest density value of the involvement entropy $\s$ around $\approx 1$, whatever countries, type of elections, etc., but also by its deviances. There are two kinds of deviances: for the fist one, $\s$ is small (which generally occurs when $\pa$ or $\pc$ is very small), for the second one, $\s$ is high (which generally results from great ratio of blank or null votes, $\pn$). We will see that these deviances are associated with a particular phenomena of civic involvement, or are simply reduced to the norm (i.e $\s\approx 1$) when the meaning of blank votes changes. 

When significantly smaller values are observed (e.g. $\s\lesssim0.85$) for cities, something appear inside towns (in average): the heterogeneity of involvement entropy over all polling stations of a given town decreases when $\s$ of the whole city decreases. In other words, considering the electorate civic involvement in a given town, the less is $\s$ for the whole town, the more the town appears homogeneous (i.e. involvement entropies, at polling station scale, over all polling stations of the town are more homogeneous between themselves). Section~E in Appendix S1 shows this point (free of statistical bias), particularly clear when the ratio $\pc$ is high (compared to cases where $\pa$ are high). This civic involvement phenomenon for towns with small $\s$ can be seen as a signature of something `new' which appears when deviance of the norm occurs. 

\begin{figure}[h!]
\centering
   \includegraphics[width=0.45\linewidth, height=4cm, clip=true]{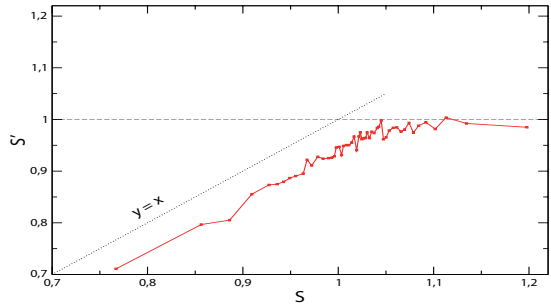}
\caption{{\bf A modified involvement entropy, $\s'$, where Blank Votes are grouped with Valid Votes, with respect to the involvement entropy $\s$}, for $\approx500$ Swiss referendums. (See Appendix S1, Section~F, for a deeper discussion.) Each point corresponds to the average of about 10 referendums. Note the plateau $\s'\approx 1$ for $\s\gtrsim 1.05$.}
\label{fblancs-modif}
\end{figure}

On the other hand, elections where significantly $\s>1$, typically corresponds to cases where there has been an appeal (from political parties, citizens blogs, etc.) to vote blank or null, which adds civic-involvement `tensions' to the election. It is remarkable that countries which make the distinction between blank votes to null votes, provide, by considering blank votes like the valid votes in favor of one of the list of choices, a modified involvement entropy $\s'\approx 1$ whenever the involvement entropy is $\s>1$. (When blank votes are grouped with votes according to the list of choices, the modified involvement entropy $\s'$ is equal to $\s(\pa,\,\pc+\pwh,\,\pnu)$ in Eq.~(\ref{es3}), and not $\s(\pa,\,\pc,\,\pwh+\pnu)$ as for the usual involvement entropy, where $\pwh$, $\pnu$ and $\pn=\pwh+\pnu$ mean respectively ratios of blank votes, null votes and blank or null votes.) See the striking plateau in Fig.~\ref{fblancs-modif} for Swiss referendums, which shows a modified involvement entropy $\s'\approx 1$ when $\s>1$. Moreover, Section~F in Appendix S1 clearly shows this point, e.g. for European Parliament elections in Italy, and for Referendums in Spain. Hence the fact that $\s>1$ boils down to a modified involvement entropy $\approx 1$, by categorizing blank votes as Valid Votes, can be seen as the recovering of the `weak law' by the decrease of civic involvement `tensions'. The fact that a deviance of the norm is naturally reduced to the norm (the involvement entropy is around 1) as soon as blank votes are grouped with `valid votes' can be seen not as an haphazardly occurrence but rather as a signature of the norm in a larger sense.\\

\begin{figure}[h!]
\centering
   \includegraphics[width=0.45\linewidth, height=6cm, clip=true]{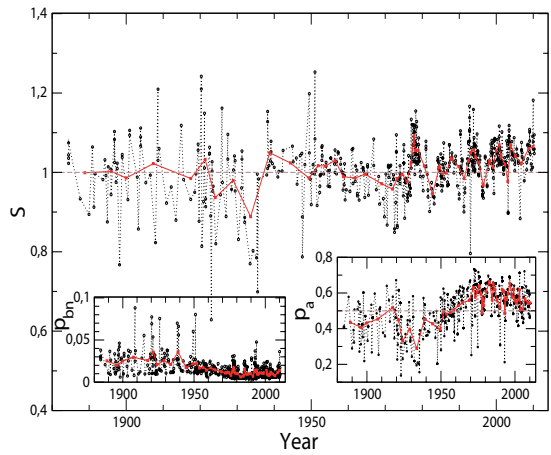}
\caption{{\bf Time evolution of the involvement entropy, $\s$, of 531 Swiss referendums}, at large scale. Each point corresponds to the average (weighted by the number of registered voters) over all Swiss {\it cantons} (25 or 26 in quantity). In red (as a guide view): average values over $\approx$10 referendums. The inset show sames things, but for ratio of abstentionists, $\pa$, and the rartio of blank and null votes, $\pn$.}
\label{fdyn-ch-R}
\end{figure}

Now let us comment about the use of the term `weak law'. In one hand, the common value $\s\approx 1$ (for the most populated municipalities in recent times) appears as a kind of law of a phenomenon not yet measured up to now. This phenomenon concerns the involvement of the electorate, from a civic point of view. A kind of law, because it occurs very frequently, with strong regularities despite wide disparities across elections. As we have seen, it implies the existence of particular correlations between $\pa$ and $\pn$. In other hand, this is clearly not a `hard' or `strong' law since noticeable deviations are observed. One cannot exclude that a `strong' law exists, encapsulating more regularities for the most populated municipalities (e.g. by taking into account not only $\pa$, $\pn$ and $\pc$, but also parameters characterizing the political context, the number of valid votes for different choices, etc.). Such more global law might explain why $\s\approx 1$ appears in recent times and why this phenomenon is not observed for small municipalities. In any case, we believe that this weak law of recent urban civic involvement shows up as a consequence of some robust electoral behavior. As one more illustration, Swiss referendums show (at the {\it canton} scale) $\s\approx 1$ with small fluctuations, and this from 1880s to nowadays (see Fig.~\ref{fdyn-ch-R}).\\

To conclude, the main finding of this work, based on the analysis of a wide number of elections from 11 different countries, is that a common stylized fact emerges: in recent elections, the distribution of the involvement entropy is found to be sharply peaked near $\s\approx1$, in high populated municipalities (and thus also at national levels). This universal property is remarkable given the wide disparities across countries (and even within countries for different elections) in political mores, voting systems, in the way that lists of registered voters are established (on a voluntary basis or automatically, etc.), and so on.

Moreover, $\s\approx1$ appears to be very stable in time whenever it occurs for one kind of election, as for example European Parliament elections in Western Europe, and particularly remarkably for the Swiss referendums since 1884. We propose to designate this strong regularity, neither a `hard law' nor a mere statistical artefact, as a `weak law' of electoral involvement characteristic of modern democracies in urban cities. We suggest that the existence of this weak law is the signature of an emerging collective behavioral norm. More studies and analysis would be necessary in order to better understand its conditions of realizations and its meaning (at the individual scale and/or at macro scale). Moreover, it should be very interesting for forthcoming studies, notably to know if this `weak law' also occurs in emergent countries, in new democratic countries, in great cities (whatever they are), etc.

The present study calls for a different point of view than those commonly used in Political Sciences. We do not work within the classical paradigm explaining the electoral behavior with sociological or ethnic even institutional or rational choice variables. Our propose is to change perspective of observation, using very large sets of data, looking for regularities -- stylized facts --, without restricting the analysis to a particular category which could be based on chronology, space, institutional or national specificities. At a `macro' level, using aggregated data, and not at the individual scale, this new view point focuses on (1) the involvement or the mobilization of the electorate, and (2) a measure of heterogeneity or, otherwise stated, of order and disorder. The question asked here to electoral data is not why a more or less rational citizen participates or not to an election, but {\it how is the degree of disorder of civic involvement of the electorate}.




\section*{Acknowledgments}
C. B. would like to thank Brigitte Hazart, from the French \textit{Minist\`ere de l'Int\'erieur, bureau des \'elections et des \'etudes politiques}; Nicola A. D'Amelio, from the Italian \textit{Ministero dell'interno, Direzione centrale dei servizi elettorali}; Radka Sm\'idov\'a, from the \textit{Czech Statistical Office, Provision of electronic outputs}; Claude Maier and Madeleine Schneider, from the Swiss \textit{Office f\'ed\'eral de la statistique, Section Politique, Culture et M\'edias}; Alejandro Vergara Torres, Antonia Ch\'avez, from the Mexican \textit{Instituto Federal Electoral}; Matthias Klumpe from the German \textit{Amt f\"ur Statistik Berlin-Brandenburg}; anonymous correspondents of the \textit{\'Elections Canada, Centre de renseignements} and of the Spanish \textit{Ministerio del Interior, Subdirecci\'on General de Pol\'itica Interior y Procesos Electorales}, for their explanations and also for the great work they did to gather and make available the electoral data that they send us.\\
C.B. is grateful to Arnaud Faure for his sociological insights, Lionel Tabourier for his help and enlightening discussions, Alexandra Reisigl for her enlightenment in the overlap idea and help for translation; and the authors thank Daniel Boy and Bruno Cautr\`es for very useful comments.


\clearpage
\section*{Figure Legends}

\clearpage
\section*{Tables}

\begin{table*}[h!]
\begin{small}
\begin{tabular*}{\hsize}{@{\extracolsep{\fill}} | l l l | l l l | l l l | l l l | l l l | l l l |}
\hline
At & 13 & 1945 & Ca* & 5 & 1945 & CH & 3 & 1884 & Cz & 1 & 1990 & Fr* & 20 & 1946 & Ge & 7 & 1949\\
It & 4 & 1946 & Mx* & 4 & 1991 & Pl* & 11 & 1990 & Ro* & 4 & 1990 & Sp & 4 & 1976 & & & \\
\hline
\end{tabular*}
\end{small}
\caption{{\bf Countries where elections are analyzed in this paper} (first column). Number of elections studied at the municipality scale (second column), and the date from which they are studied at national or provincial scale (third column) -- even if it is before the end of the compulsory voting in Austria and in Italy. Star indicates that electoral data are also known at polling station level. Number of municipalities per country: $\approx 2400$ in Austria (At); $\approx 7700$ in Canada (Ca); $\approx 2700$ in Switzerland (CH); $\approx 6400$ in Czech Republic (Cz); $\approx 36000$ in Metropolitan France (Fr); $\approx 12000$ in Germany (Ge); $\approx 8100$ in Italy (It); $\approx 2400$ in Mexico (Mx); $\approx 2500$ in Poland (Pl); $\approx 3200$ in Romania (Ro); $\approx 8100$ in Spain (Sp). See Appendix S1, Section~A, for more details.}
\label{tdataset}
\end{table*}

\begin{table*}[h!]
\begin{tabular*}{\hsize}{@{\extracolsep{\fill}} | l l l r r c c c c|}
\hline
Ctry & date & $n_{el}$ & $N_{min}$ & $\overline{N}$ & $N_{bin}/N_{Ctry}$ & $\smoy < 0.98$ & $\smoy\in[0.98,\,1.08]$ & $1.08 < \smoy$\\
\hline
 & $t < 2000$ & & & & & & & \\
Fr & & 8 & 32000 & 69000 & 18$\%$ & 1 & 2 & 5\\
At & & 6 & 7000 & 26000 & 44$\%$ & 3 & 3 & 0\\
Ca & & 1 & 30000 & 94000 & 48$\%$ & 0 & 1 & 0\\
\hline
 & $t \geq 2000$ & & & & & & & \\
Fr & & 12 & 33000 & 70000 & 17$\%$ & 3 & 8 & 1\\
At & & 7 & 7000 & 27000 & 43$\%$ & 3 & 3 & 1\\
Pl & & 11 & 39000 & 120000 & 39$\%$ & 2 & 8 & 1\\
Ge & & 7 & 68000 & 190000 & 30$\%$ & 3 & 4 & 0\\
Ca & & 4 & 53000 & 83000 & 38$\%$ & 0 & 4 & 0\\
It & & 4 & 48000 & 150000 & 31$\%$ & 2 & 0 & 2\\
Sp & & 4 & 48000 & 160000 & 47$\%$ & 2 & 2 & 0\\
Mx & & 4 & 130000 & 370000 & 53$\%$ & 0 & 3 & 1\\
Ro & & 4 & 20000 & 87000 & 47$\%$ & 1 & 2 & 1\\
CH & & 3 & 7500 & 18000 & 37$\%$ & 0 & 3 & 0\\
Cz & & 1 & 14000 & 36000 & 43$\%$ & 0 & 1 & 0\\
\hline
\end{tabular*}
\caption{{\bf Basic information about the bin of the $\approx$100 most populated municipalities} per country (Ctry). $n_{el}$ means the number of elections analyzed. The municipality of this bin with the lowest number of registered voters is written as $N_{min}$; the average value of $N$ over these municipalities, as $\overline{N}$; and the ratio of registered voters which belongs to this bin over those in the whole country, as $N_{bin}/N_{Ctry}$. $\smoy$ is classified according to values $0.98$ and $1.08$.}
\label{tdern}
\end{table*}


\clearpage
\begin{center}\begin{Large}{\bf Between order and disorder:\\a `weak law' on recent electoral behavior among urban voters?}\vspace{0.5cm}\\
Christian Borghesi, Jean Chiche and Jean-Pierre Nadal\end{Large}\end{center}%

\renewcommand{\theequation}{S\arabic{equation}}
\setcounter{equation}{0}  
\renewcommand{\thefigure}{S\arabic{figure}}
\setcounter{figure}{0}
\renewcommand{\thetable}{S\arabic{table}}
\setcounter{table}{0}
\section*{Supporting Information: Appendix S1}
\label{ssupplement}

\subsection*{A. Data}

\noindent
$\bullet$ {\bf Elections studied at municipality scale}\\
Table~\ref{si-tdata-mun} gives more details about the 76 elections studied in this paper at the municipality scale. There are: 13 elections from Austria~\cite{data-mun-austria} ($\approx 2400$ municipalities)~\footnote{Corrections due to \textit{wahlkarten} or postal votes are taking account from the national level, i.e. in this paper, each municipality receive from voting cards a number of votes and valid votes proportional to its number of population, and at the same ratio for every municipality.}; 5 from Canada~\cite{data-mun-ca} ($\approx 7700$ municipalities); 1 from Czech Republic~\cite{data-mun-cz} ($\approx 6400$ municipalities); 20 from Metropolitan France~\cite{data-mun-fr} ($\approx 36000$ municipalities); 7 from Germany~\cite{data-mun-ge} ($\approx 12000$ municipalities)~\footnote{Chamber of Deputies (D) elections refer to the {\it German Bundestag} elections. \textit{Land} Parliament elections at time less or equal to 2004 (or 2010) in each {\it Land} are written here as `2004 Ld' (or `2010 Ld'). Postal votes ({\it briehwahlen}) are usually taken account at {\it Landkreis} scale (they are distributed in municipalities, according to their populations), when it is it possible to do it. Nevertheless, these corrections provide a very small difference in Fig.~5, especially for high population-size bins.}; 4 from Italy~\cite{data-mun-it} ($\approx 8100$ municipalities); 4 from Mexico~\cite{data-mun-mx} ($\approx 2400$ municipalities)~\footnote{The 2006 \textit{Senador} election (not studied here) gives a very near statistics of $\s$ than the (P) and (D) elections that also occur at the same time.}; 11 from Poland~\cite{data-mun-pl} ($\approx 2500$ municipalities)~\footnote{The Chamber of Deputies (D) election is the \textit{Sejm} Chamber election.}; 4 from Romania~\cite{data-mun-ro} ($\approx 3200$ municipalities)~\footnote{The referendum studied here is about \textit{the reduction of the number of parliamentarians to a number of 300 persons}, and not about the \textit{adoption of a unicameral Parliament} held on the same time. The latter one is not known at the polling station level.} \footnote{Some Romanian electors, not registered in the \textit{lista electorala permanenta}, are able to vote. For this country, we pursue to write $N$ the Number of Register Voters, $\nv$ the registered electors who take part to the election, and $\nn$ the number of Null and Blank Votes that the Registered Voters could make (even if the latter data is not known.) Romanian electoral data gather for each municipality, $N$, $\nv$, $\nv(tot)$ (the total number of votes), and $\nn(tot)$ the total number of Null and Blank votes. Assuming that registered electors and not registered electors vote Null and Blank in the same way (i.e. $\frac{\nn(tot)}{\nv(tot)}=\frac{\nn}{\nv}$), we deduce $\nn$.}, 4 from Spain~\cite{data-mun-sp} ($\approx 8100$ municipalities) and 3 from Switzerland~\cite{data-mun-ch} ($\approx 2700$ municipalities)~\footnote{The referendums or \textit{votations} `R(a)' and `R(b)' respectively occurred the 11 of March and the 17 of June. The Legislative (D) election refers to the \textit{Conseil National} election.}. 

Table~\ref{si-tdata-mun} also gives basic statistics over the $\approx 100$ ($\approx 200$ for France) most populated municipalities.\\

\begin{table*}
\begin{footnotesize} 
\begin{tabular*}{\hsize}{@{\extracolsep{\fill}} | l c c c c | l c c c c |}
\hline
Id & $\smoy$ & $\overline{\pa}$ & $\overline{\pc}$ & $\overline{\pn}$\footnotesize{($\overline{\pwh}$)} & Id & $\smoy$ & $\overline{\pa}$ & $\overline{\pc}$ & $\overline{\pn}$\footnotesize{($\overline{\pwh}$)}\rule[-7pt]{0pt}{18pt}\\ 
\hline
Fr 1992 R & {\bf 1.02}$\pm$0.04 & 0.32 & 0.66 & 0.018 & Fr 1993 D & 1.09$\pm$0.04 & 0.34 & 0.63 & 0.028\\
Fr 1994 E & 1.12$\pm$0.03 & 0.48 & 0.50 & 0.020 & Fr 1995 P1 & 0.91$\pm$0.04 & 0.24 & 0.74 & 0.018\\
Fr 1995 P2 & {\bf 1.01}$\pm$0.07 & 0.23 & 0.73 & 0.044 & Fr 1997 D & {\bf 1.08}$\pm$0.03 & 0.36 & 0.62 & 0.024\\
Fr 1998 rg & 1.11$\pm$0.03 & 0.46 & 0.52 & 0.019 & Fr 1999 E & 1.11$\pm$0.03 & 0.54 & 0.44 & 0.020\\
Fr 2000 R & {\bf 1.02}$\pm$0.07 & 0.71 & 0.25 & 0.036 & Fr 2002 P1 & {\bf 1.01}$\pm$0.04 & 0.31 & 0.67 & 0.019\\
Fr 2002 P2 & 0.95$\pm$0.07 & 0.21 & 0.75 & 0.035 & Fr 2002 D & {\bf 1.02}$\pm$0.04 & 0.37 & 0.62 & 0.010\\
Fr 2004 rg & 1.10$\pm$0.04 & 0.41 & 0.57 & 0.021 & Fr 2004 E & {\bf 1.04}$\pm$0.03 & 0.57 & 0.42 & 0.010\\
Fr 2005 R & {\bf 1.00}$\pm$0.05 & 0.32 & 0.66 & 0.014 & Fr 2007 P1 & 0.72$\pm$0.08 & 0.17 & 0.82 & 0.010\\
Fr 2007 P2 & 0.84$\pm$0.06 & 0.17 & 0.80 & 0.032 & Fr 2007 D & {\bf 1.04}$\pm$0.03 & 0.42 & 0.57 & 0.009\\
Fr 2009 E & {\bf 1.03}$\pm$0.05 & 0.60 & 0.39 & 0.012 & Fr 2010 rg & {\bf 1.06}$\pm$0.03 & 0.57 & 0.42 & 0.012\\
\hline
At 1994 D & 0.81$\pm$0.11 & 0.20 & 0.78 & 0.016 & At 1995 D & 0.73$\pm$0.10 & 0.15 & 0.83 & 0.018\\
At 1996 E & {\bf 1.04}$\pm$0.04 & 0.33 & 0.65 & 0.021 & At 1998 P & {\bf 1.00}$\pm$0.10 & 0.27 & 0.70 & 0.032\\
At 1999 E & {\bf 1.06}$\pm$0.05 & 0.52 & 0.46 & 0.013 & At 1999 D & 0.82$\pm$0.09 & 0.22 & 0.77 & 0.011\\
At 2002 D & 0.73$\pm$0.10 & 0.17 & 0.81 & 0.011 & At 2004 P & {\bf 1.04}$\pm$0.09 & 0.31 & 0.66 & 0.028\\
At 2004 E & {\bf 1.03}$\pm$0.05 & 0.59 & 0.40 & 0.010 & At 2006 D & 0.87$\pm$0.09 & 0.24 & 0.74 & 0.012\\
At 2008 D & 0.88$\pm$0.08 & 0.24 & 0.75 & 0.014 & At 2009 E & {\bf 1.04}$\pm$0.04 & 0.55 & 0.44 & 0.009\\
At 2010 P & 1.16$\pm$0.06 & 0.48 & 0.49 & 0.034 & & & & & \\
\hline
Pl 2000 P1 & {\bf 0.98}$\pm$0.03 & 0.36 & 0.63 & 0.006 & Pl 2001 D & 1.09$\pm$0.02 & 0.52 & 0.46 & 0.015\\
Pl 2003 R & {\bf 0.98}$\pm$0.02 & 0.37 & 0.62 & 0.004 & Pl 2004 E & 0.79$\pm$0.07 & 0.78 & 0.22 & 0.005\\
Pl 2005 D & {\bf 1.06}$\pm$0.03 & 0.58 & 0.41 & 0.013 & Pl 2005 P1 & {\bf 1.02}$\pm$0.01 & 0.49 & 0.51 & 0.003\\
Pl 2005 P2 & {\bf 1.03}$\pm$0.01 & 0.47 & 0.53 & 0.006 & Pl 2007 D & {\bf 1.05}$\pm$0.03 & 0.42 & 0.57 & 0.010\\
Pl 2009 E & 0.87$\pm$0.06 & 0.73 & 0.27 & 0.004 & Pl 2010 P1 & {\bf 1.01}$\pm$0.02 & 0.43 & 0.57 & 0.004\\
Pl 2010 P2 & {\bf 1.03}$\pm$0.02 & 0.43 & 0.56 & 0.007 & & & & &\\
\hline
Ge 2002 D & 0.83$\pm$0.07 & 0.22 & 0.77 & 0.009 & Ge 2004 Ld & {\bf 1.02}$\pm$0.04 & 0.41 & 0.58 & 0.007\\
Ge 2004 E & {\bf 1.02}$\pm$0.05 & 0.59 & 0.40 & 0.009 & Ge 2005 D & 0.87$\pm$0.06 & 0.24 & 0.75 & 0.011\\
Ge 2009 E & {\bf 1.00}$\pm$0.05 & 0.60 & 0.40 & 0.006 & Ge 2009 D & 0.95$\pm$0.05 & 0.30 & 0.69 & 0.009\\
Ge 2010 Ld & {\bf 1.04}$\pm$0.03 & 0.43 & 0.56 & 0.009 & & & & & \\
\hline
Ca 1997 D & {\bf 1.00}$\pm$0.04 & 0.37 & 0.62 & 0.009 & Ca 2000 D & {\bf 1.03}$\pm$0.03 & 0.44 & 0.56 & 0.006\\
Ca 2004 D & {\bf 1.02}$\pm$0.02 & 0.46 & 0.54 & 0.004 & Ca 2006 D & {\bf 1.01}$\pm$0.02 & 0.44 & 0.56 & 0.003\\
Ca 2008 D & {\bf 1.02}$\pm$0.02 & 0.49 & 0.51 & 0.003 & & & & & \\
\hline
It 2004 E & 1.11$\pm$0.12 & 0.29 & 0.66 & 0.053\footnotesize{(0.023)} & It 2006 D & 0.78$\pm$0.13 & 0.17 & 0.81 & 0.020\footnotesize{(0.007)}\\
It 2008 D & 0.89$\pm$0.12 & 0.20 & 0.77 & 0.027\footnotesize{(0.008)} & It 2009 E & 1.08$\pm$0.10 & 0.36 & 0.61 & 0.034\footnotesize{(0.013)}\\
\hline
Mx 2003 D & {\bf 1.04}$\pm$0.05 & 0.59 & 0.40 & 0.013 & Mx 2006 D & {\bf 1.04}$\pm$0.04 & 0.40 & 0.58 & 0.012\\
Mx 2006 P & {\bf 1.03}$\pm$0.04 & 0.40 & 0.59 & 0.010 & Mx 2009 D & 1.11$\pm$0.06 & 0.56 & 0.41 & 0.027\\
\hline
Ro 2009 E & 0.73$\pm$0.09 & 0.81 & 0.18 & 0.008 & Ro 2009 R & 1.09$\pm$0.02 & 0.55 & 0.44 & 0.017\\
Ro 2009 P1 & {\bf 1.05}$\pm$0.02 & 0.52 & 0.48 & 0.008 & Ro 2009 P2 & {\bf 1.04}$\pm$0.02 & 0.50 & 0.50 & 0.006\\
\hline
Sp 2004 D & 0.92$\pm$0.07 & 0.24 & 0.74 & 0.020\footnotesize{(0.014)}& Sp 2004 E & {\bf 1.01}$\pm$0.06 & 0.57 & 0.42 & 0.006\footnotesize{(0.003)}\\
Sp 2008 D & 0.91$\pm$0.08 & 0.26 & 0.73 & 0.013\footnotesize{(0.009)}& Sp 2009 E & {\bf 1.03}$\pm$0.04 & 0.56 & 0.43 & 0.009\footnotesize{(0.006)}\\
\hline
CH 2007 R(a) & {\bf 1.04}$\pm$0.04 & 0.53 & 0.46 & 0.008\footnotesize{(0.004)}& CH 2007 R(b) & {\bf 0.99}$\pm$0.06 & 0.62 & 0.37 & 0.007\footnotesize{(0.004)}\\
CH 2007 D & {\bf 1.04}$\pm$0.05 & 0.53 & 0.47 & 0.009\footnotesize{(0.002)} & & & & & \\
\hline
Cz 2003 R & {\bf 1.07}$\pm$0.01 & 0.47 & 0.52 & 0.012 & & & & & \\
\hline
\end{tabular*}
\end{footnotesize}
\caption{{\bf Elections studied in this paper at the municipality scale.} An election is identified (Id) by its country, its year date and its nature. D: Chamber of Deputies election; E: European parliament election; P: presidential election (according to the constitution of the country, in only one round); P1 and P2: first and second round of a Presidential election; R: Referendum; Ld: German \textit{L\"ander} elections; rg: French \textit{R\'egionales} elections. For each country elections are given in a chronological order (but the 2006 Mexican Presidential (P) and Deputies (D) elections occurred the same day, and also for the 2009 Romanian Presidential (P1) and Referendum (R) elections). Even if an election needs two rounds, only the first one is considered (e.g. the French Deputies (D) and \textit{R\'egionales} (rg) elections) unless the contrary is indicated (e.g. P1 and P2). Mean values of $\s$, $\pa$, $\pc$, $\pn$(and ($\pwh$) if Blank Vote are distinguished between Null Vote), and also standard deviation only for $\s$, are given over the bin of the $\approx 100$ (or $\approx 200$ for France only) most populated municipalities. In bold text, $\smoy\in[0.98;1.08]$.}
\label{si-tdata-mun}
\end{table*}

\noindent
$\bullet$ {\bf Time evolution at the national or provincial scale}\\
The study of time evolution of $\s$ is done for the same countries as in Tab.~\ref{si-tdata-mun} and for all national elections for which we have enough data. For Austria~\cite{data-dyn-austria}, the study considers data since 1945, even if compulsory voting was abolished in the whole country in 1992 for National Council elections (D), and after 2004 for Presidential elections (P) (but in 1982 some provinces had yet done it); for Canada~\cite{data-dyn-ca}, since 1945; for Czech Republic~\cite{data-dyn-cz}, since 1990~\footnote{The 1990 and 1992 Deputies (D) elections only refer to the Parliamentary Chamber of People election. The Parliamentary Chamber of Nations and the Parliamentary National Council elections, that occurred at the same day as the previous ones, also gave approximately the same $\s$ value.}; for France~\cite{data-dyn-fr}, since 1945~\footnote{All French electoral data are from metropolitan France. Some referendums are not known at the {\it d\'epartement} scale. In these cases, $\s$ is evaluated at the national scale.}; for Germany~\cite{data-dyn-ge}, since 1949; for Italy~\cite{data-dyn-it}, since 1945 even if there were compulsory voting until 1993~\footnote{We consider the only first question asked to electors in referendums.}; for Mexico~\cite{data-dyn-mx}, since 1991; for Poland~\cite{data-dyn-pl}, since 1990~\footnote{We have not data from the 1989 Chamber of Deputies (Sejm) election nor the two referendums in 1996.}; for Romania~\cite{data-dyn-ro}, since 1990; for Spain~\cite{data-dyn-sp}, since 1976; for Switzerland~\cite{data-dyn-ch}, since 1884 for referendums (R) and since 1919 for legislative elections (D). If an election needs two rounds, the first one is considered, unless the contrary is indicated. The Mexican, Polish and Romanian Senate elections are not shown here because they occur at the same time as Chamber of Deputies elections and have very similar $\s$ results. 

Table~\ref{si-tdata-dyn} summarizes the nature of elections studied in this paper, and also the scale of aggregate data per country. Note that the last election analyzed in this paper is the Referendum which held in Italy on June 2011.~\footnote{Official results (which took into account registered voters) of the Canadian Chamber of Deputies election, held on May 2011, were not published at the time we first submitted this paper. In Fig.~\ref{si-fdyn}, the involvement entropy over all provinces would be $\approx 1.00\pm 0.02$ and respectively $0.99$ and $1.02$ for Ontario and Quebec.}\\

\begin{table}
\centering
\begin{tabular}{| l l l |}
\hline
Country & Kind of elections & Scale of aggregate data\\
\hline
At & D, E, P, R & National\\
Ca & D & Province (5-13)\\
CH & D, R & \textit{Canton} (25-26)\\
Cz & D, E, R, rg, S1, S2 & National\\
Fr & Cant, D, E, P1, P2, R, rg & \textit{d\'epartement} (90-96)\\
Ge & D, E & \textit{Land} (9-16)\\
It & D, E, R, S & National\\
Mx & D, P & National\\
Pl & D, E, P1, P2 & National\\
Ro & D, E, P1, P2, R & National\\
Sp & D, E, R & \textit{Comunidad aut\'onoma} (17-19)\\
\hline
\end{tabular}
\caption{{\bf Elections studied in this paper at large scale for their evolution in time.} Notation is the same as in Tab.~\ref{si-tdata-mun}. For Czech Republic, ``rg'' means Election into regional councils, ``S1'' and ``S2'' are respectively the first and second round of the Senate elections; for France, ``Cant'' refers to the \textit{Cantonales} elections and some referendums are only known at the national scale; for Italy, ``S'' means Senate elections, and occur at the same time as Deputies elections (D) but with older registered voters. In parenthesis, the total number of different provinces (or \textit{Cantons}, etc.), which can change in time, in the whole country.}
\label{si-tdata-dyn}
\end{table}

Websites given in the References were accessed in December 2011. Part of the database used in this paper can also be directly downloaded from~\cite{si-free-data}.\\

\noindent
$\bullet$ {\bf Elections studied at polling station scale}\\
Polling stations analysis is restricted to polling stations which belong to one of the 100 most populated municipalities (for the considered election). 31 elections at the polling station scale are studied in this paper: 5 for Canada (each Canadian election of Tab.~\ref{si-tdata-mun}), with around 25000 polling stations; 13 for France (French elections of Tab.~\ref{si-tdata-mun} since 1999), with around 7000 polling stations; 4 for Mexico (each Mexican election of Tab.~\ref{si-tdata-mun}), with around 55000 polling stations or ballot box; 5 for Poland (Polish election of Tab.~\ref{si-tdata-mun} from 2003 up to 2005), with around 8000 polling stations; and 4 for Romania (each Romanian election of Tab.~\ref{si-tdata-mun}), with around 6000 polling stations. See Tab.~\ref{si-tdata-bvot} for some basic statistics over polling stations of the 100 most populated municipalities.

\begin{table*}
\begin{tabular*}{\hsize}{@{\extracolsep{\fill}}| l c c | l c c |}
\hline
Id & $\s$ & $\tau_3$ & Id & $\s$ & $\tau_3$\\
\hline
Fr 1999 E & 1.09 $\pm$ 0.05 & -3.7 $\pm$ 0.6 & Fr 2000 R & 1.00 $\pm$ 0.11 & -4.2 $\pm$ 0.6\\
Fr 2002 P1 & 1.00 $\pm$ 0.06 & -2.1 $\pm$ 0.6 & Fr 2002 P2 & 0.93 $\pm$ 0.10 & -0.7 $\pm$ 0.6\\
Fr 2002 D & 1.01 $\pm$ 0.06 & -3.2 $\pm$ 0.7 & Fr 2004 rg & 1.09 $\pm$ 0.05 & -2.8 $\pm$ 0.6\\
Fr 2004 E & 1.03 $\pm$ 0.05 & -4.6 $\pm$ 0.7 & Fr 2005 R & 0.99 $\pm$ 0.07 & -2.5 $\pm$ 0.7\\
Fr 2007 P1 & 0.71 $\pm$ 0.11 & -1.3 $\pm$ 0.7 & Fr 2007 P2 & 0.83 $\pm$ 0.09 & -0.1 $\pm$ 0.7\\
Fr 2007 D & 1.03 $\pm$ 0.04 & -3.8 $\pm$ 0.7 & Fr 2009 E & 1.02 $\pm$ 0.07 & -4.6 $\pm$ 0.7\\
Fr 2010 rg & 1.04 $\pm$ 0.05 & -4.4 $\pm$ 0.7 & & & \\
\hline
Ca 1997 D & 0.98 $\pm$ 0.08 & -3.3 $\pm$ 1.3 & Ca 2000 D & 1.00 $\pm$ 0.06 & -4.1 $\pm$ 1.1\\
Ca 2004 D & 1.00 $\pm$ 0.05 & -4.4 $\pm$ 0.9 & Ca 2006 D & 0.99 $\pm$ 0.05 & -4.4 $\pm$ 0.8\\
Ca 2008 D & 1.00 $\pm$ 0.05 & -4.6 $\pm$ 0.9 & & & \\
\hline
Pl 2003 R & 0.95 $\pm$ 0.10 & -4.0 $\pm$ 0.9 & Pl 2004 E & 0.83 $\pm$ 0.13 & -6.3 $\pm$ 0.8\\
Pl 2005 D & 1.05 $\pm$ 0.08 & -4.1 $\pm$ 0.9 & Pl 2005 P1 & 1.00 $\pm$ 0.07 & -4.9 $\pm$ 0.9\\
Pl 2005 P2 & 1.01 $\pm$ 0.05 & -4.1 $\pm$ 0.9 & & & \\
\hline
Mx 2003 D & 1.03 $\pm$ 0.07 & -4.3 $\pm$ 0.9 & Mx 2006 D & 1.02 $\pm$ 0.07 & -3.2 $\pm$ 0.8\\
Mx 2006 P & 1.01 $\pm$ 0.07 & -3.4 $\pm$ 0.8 & Mx 2009 D & 1.11 $\pm$ 0.10 & -3.5 $\pm$ 0.9\\
\hline
Ro 2009 E & 0.70 $\pm$ 0.13 & -6.6 $\pm$ 0.9 & Ro 2009 R & 1.08 $\pm$ 0.05 & -3.8 $\pm$ 0.7\\
Ro 2009 P1 & 1.04 $\pm$ 0.03 & -4.5 $\pm$ 0.7 & Ro 2009 P2 & 1.04 $\pm$ 0.03 & -4.4 $\pm$ 0.7\\
\hline
\end{tabular*}
\caption{{\bf Elections studied at the polling station level.} An election is identified (Id) by its country, its year date and its nature. Mean value and standard deviation of $\s$ and of $\tau_3$ (see the SI Section~D) over ballot boxes in the 100 most populated municipalities.}
\label{si-tdata-bvot}
\end{table*}

\clearpage
\subsection*{B. More details on data analysis}%

\begin{figure*}[h!]
\centering
\includegraphics[width=16cm, height=10.5cm, clip=true]{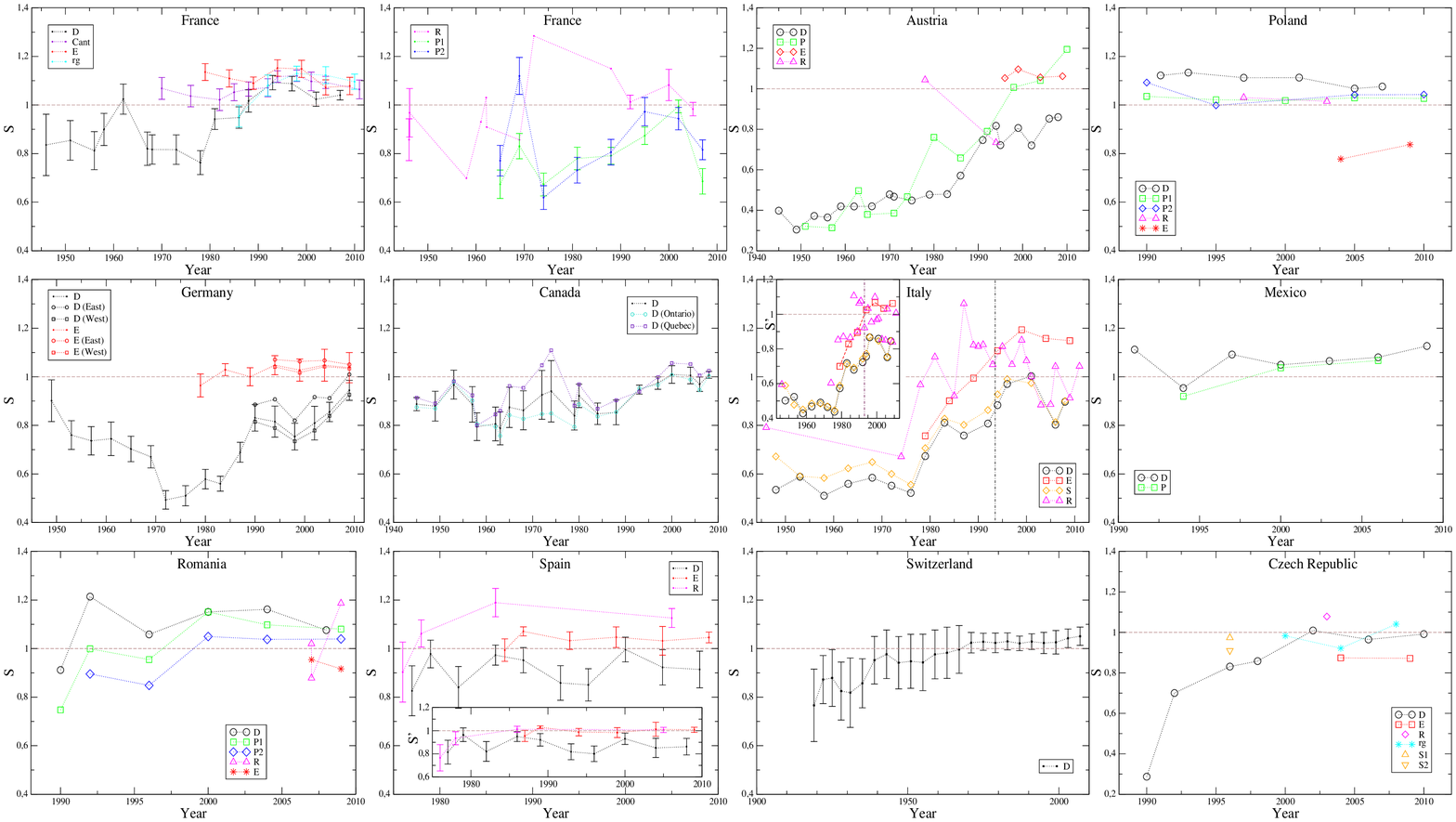}
\caption{{\bf Time evolution of the mean involvement entropy at large scale} (national, provincial, etc.). See Section~A and Tab.~\ref{si-tdata-dyn}, for more details and also for the end of compulsory voting in Italy (cf. vertical dashed line) and in Austria. Whenever the scale of aggregate data is lower than the national one, standard-deviations (weighted by the number of registered voters) are also shown as error bars. Italian and Spanish graph insets show a variant of $\s$ where Blank Votes are categorized as Valid Votes (see Section~F for more discussion). See text for more explanation about some French curves.}
\label{si-fdyn}
\end{figure*}

Fig.~\ref{si-fdyn} gathers all the available data (see in the SI, Section~A for more details) at a large aggregate scale (country, province, {\it d\'epartement}, etc.). When the scale of aggregate data is lower than the national one, each point corresponds to a weighted (by population-size) mean value of involvement entropies at lower scale (province, {\it d\'epartement}, etc.), and standard deviation is also given as error bar. The cases where Blank Votes are distinguished from Null Votes (i.e. in Italy, Spain and Switzerland), call for a specific discussion (see the SI, Section~F).

Let us comment Fig.~\ref{si-fdyn} on the case of the Chamber of Deputies elections in France, at the large scale called {\it d\'epartement} (96 in quantity for metropolitan France, actually). One sees an involvement entropy frequently equal to $\approx0.8$ until 1981, which then increases and gets greater than $1$ until 2000, and decreases a little and stabilizes to $\s\approx1$after 2000. So, the civic involvement of the electorate (at the {\it d\'epartement} scale) is relatively ordered until 1981 and get more and more disordered until 2000. After 2000, $\s$ seems to stabilize to a common value $\s\approx1$ which is also reached for the European Parliament elections and for local elections at different scales, such as the {\it R\'egionales} ($\sim$ states) and the {\it Cantonales} ($\sim$ counties) elections.

\begin{figure*}[h!]
\centering 
\includegraphics[width=16cm, height=6cm, clip=true]{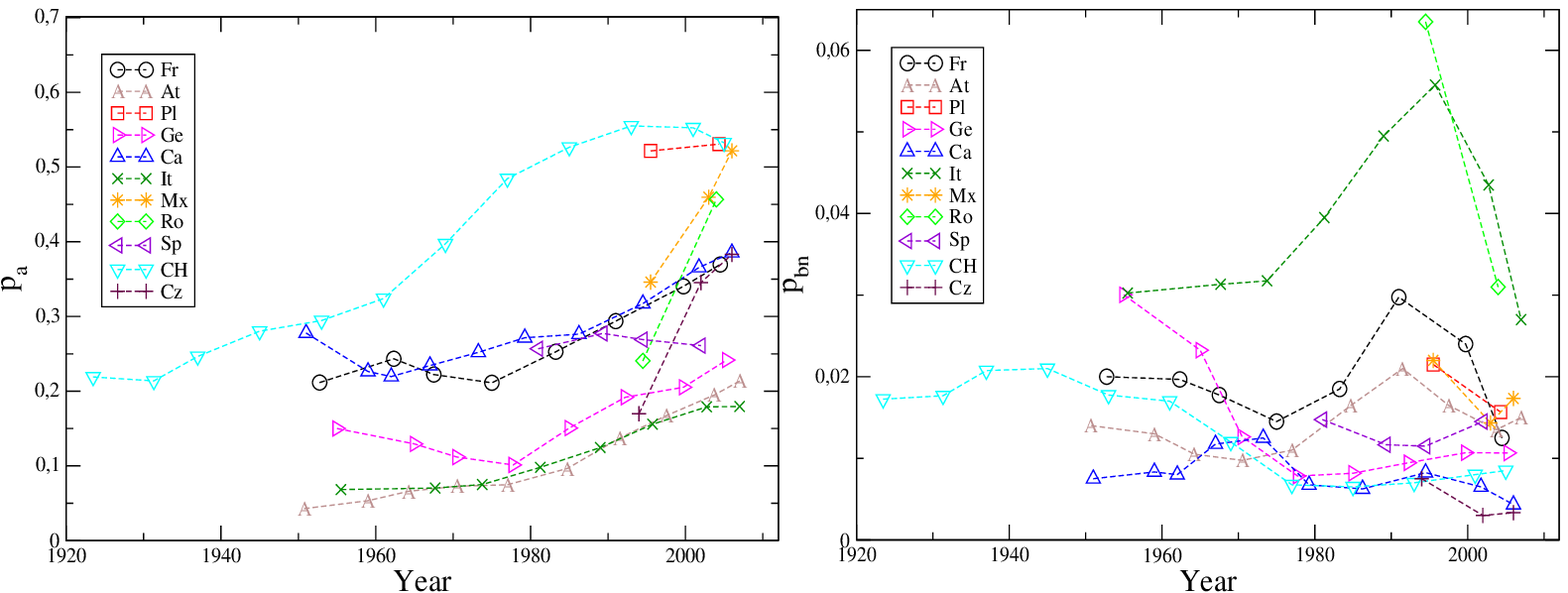}
\caption{{\bf Moving average, as a function of time, per country of $\pa$ and $\pn$ at national scale} for Chamber of Deputies elections. The average is made over 4 elections. Left: about ratio of registered voters who do not take part to the election ($\pa$); Right: about Blank and Null ratio ($\pn$).}
\label{fp-dyn}
\end{figure*}

\begin{figure*}
\centering
\includegraphics[width=16cm, height=6cm, clip=true]{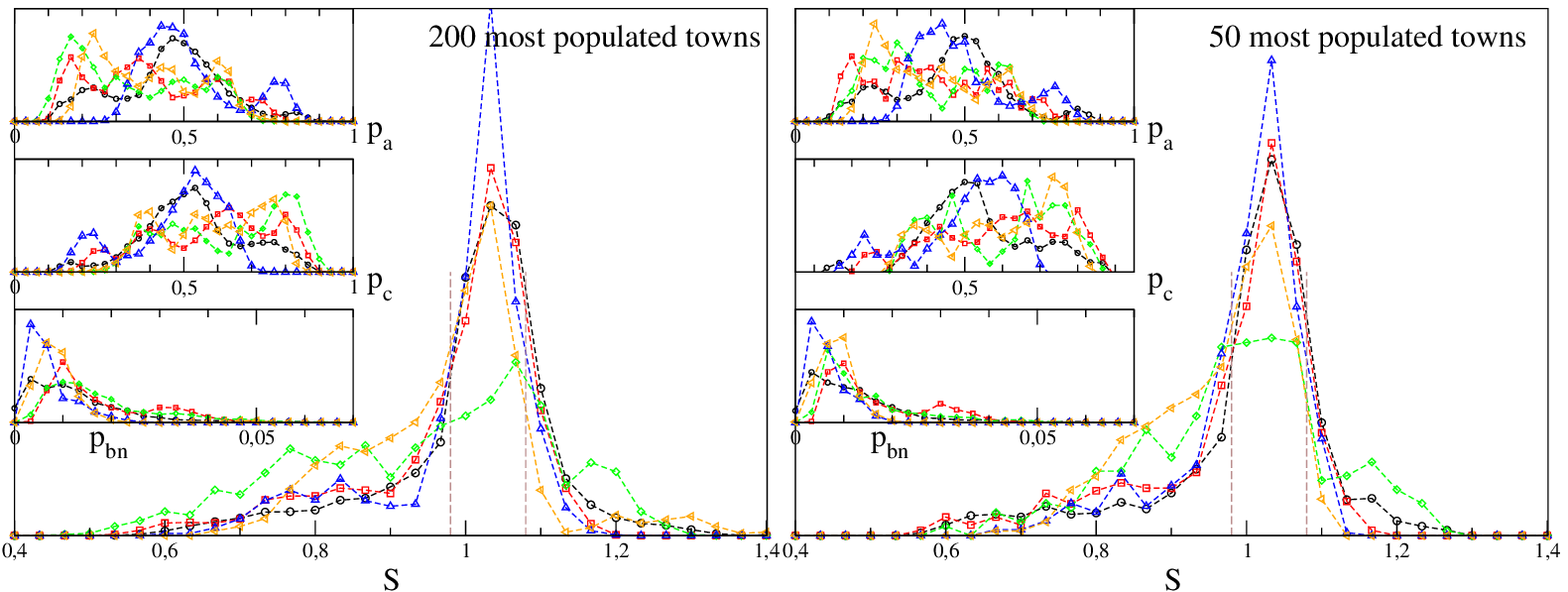}
\caption{{\bf Histograms of $\s$ for the $\approx$ 200 (left) and 50 (right) most populated municipalities}, similarly to Fig.~8-d (with 100 most populated municipalities for the latter one).}
\label{si-fhisto-size}
\end{figure*}


\begin{figure*}[h!]
\centering
\includegraphics[width=16cm, height=6cm, clip=true]{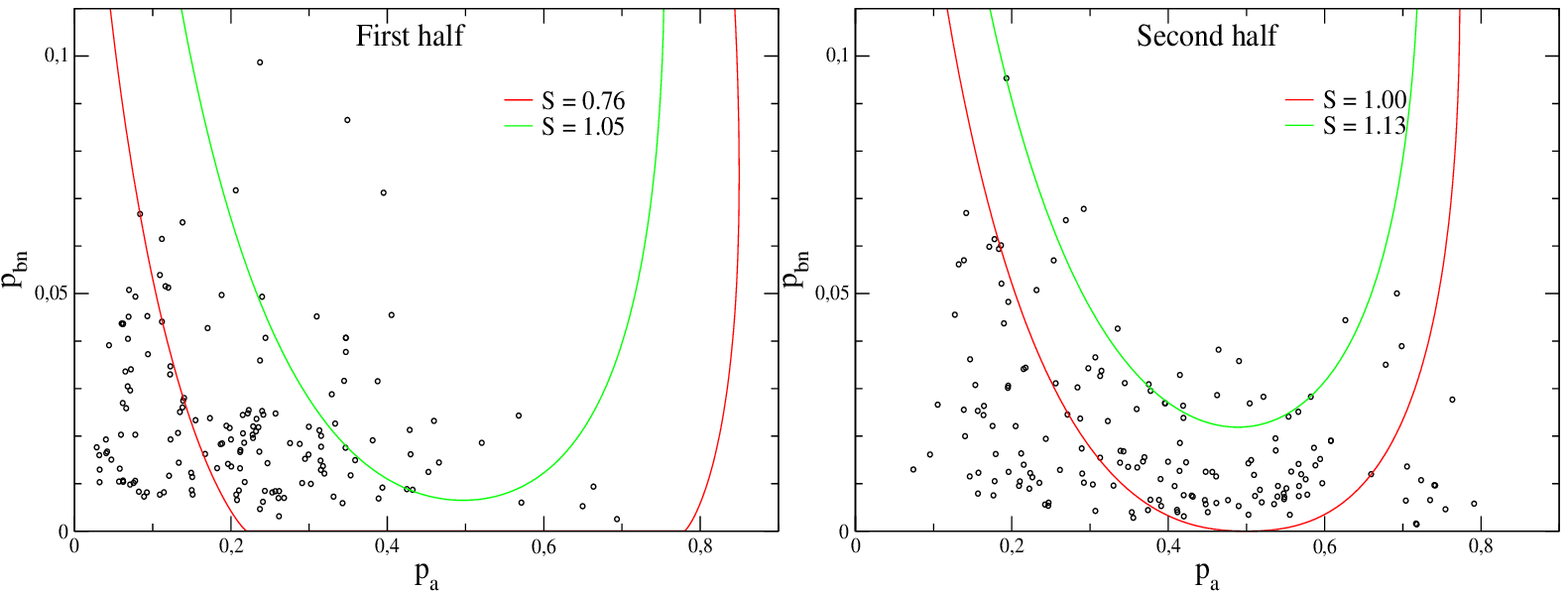}
\caption{{\bf Evolution in time of scatter plots of $(\pa,\,\pn)$ at national level} of 321 elections. Elections are divided into the two groups in the same manner as in Fig.~6. Curves give the sets of points $(\pa,\,\pn)$ such that $\s(\pa, \pn)$ is equal to one of the two endpoints of the minimal interval of $\s$ which contains $50\%$ of events. Note if $\s$ is equal to the average value (weighted by the population size) at lower aggregate scale (as provinces, {\it d\'epartements}, etc.) like in Fig.~6, the peak of $\s$ near $\s\approx 1$ would be more narrowed and more centered on $\s=1$}
\label{fpapb-scatter-dyn}
\end{figure*}

\begin{figure*}[h!]
\centering
\includegraphics[width=16cm, height=6cm, clip=true]{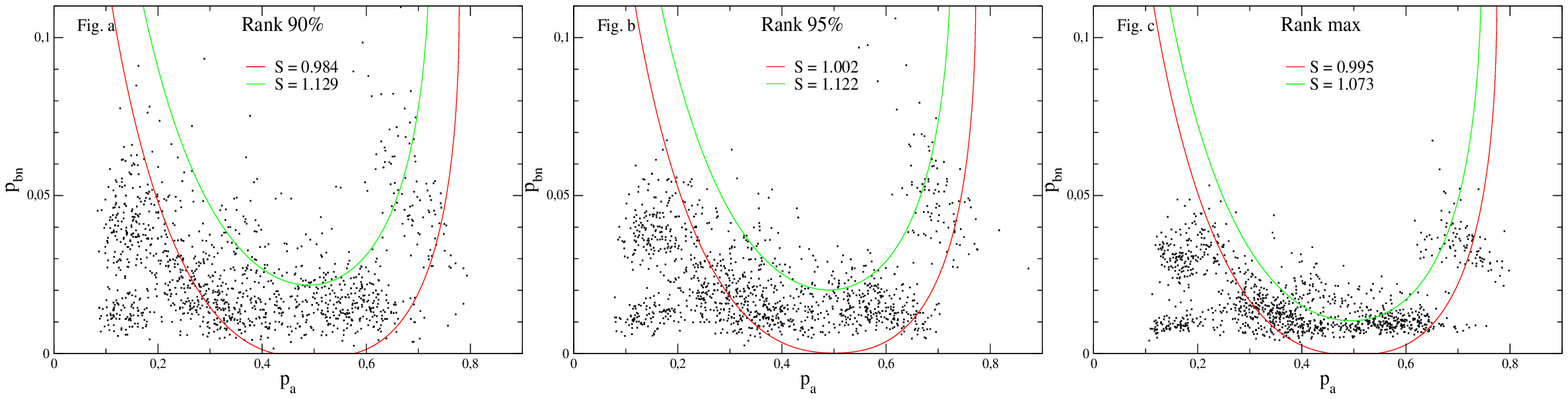}
\caption{{\bf Scatter plots of $(\pa,\,\pn)$ of French municipalities according to their relative population size}, over elections since 2000 (similarly as in Fig.~8-b, c, d). The sets of points $(\pa,\,\pn)$ such that $\s(\pa, \pn)$ is equal to one of the two endpoints of the minimal interval of $\s$ which contains $50\%$ of events (as in Fig.~9 for the most populated municipalities) are also plotted.}
\label{fpapb-scatter}
\end{figure*}

\clearpage
\subsection*{C. Finite size effects}%
We show in this section that finite size effects over municipality-size, $N$, on the entropy-involvement $\s$, are relatively small for the most populated municipalities. Biases due to finite size effects have two possible origins: (1) level of aggregation of the data, over $N$ about a hundred to a million, influences $\s$ measures, and (2) a statistical effect due to large numbers. Without a loss of generality, we examine these two biases for French electoral data -- with 20 elections at the municipality scale and $13$ at the polling station level, cf. the SI Section~A. Lastly we show that the distribution of the involvement entropy which is sharply peaked near $\s\approx 1$ for most populated towns is not due to considering a large number of $N$ per town.\\

{\bf (1) Scale at which data are aggregated}

French municipality sizes range from around $10$ to around $100,000$. In order to investigate how aggregate data scale modifies the measurement of the involvement entropy $\s$, for each municipality we compare the results at the municipality scale with the one done at the polling station scale. Registered voters per polling station do not exceed around one thousand in France. We compare for a municipality its involvement entropy, $\s$, measured at the municipality level, to the mean value, $\sbvot$, of the involvement entropy over all the polling stations in the the considered municipality. Convexity of the logarithmic function implies that the later is at most equal to the former.
For each of the $200$ most populated French municipalities, and for each of the $13$ French elections known at the polling station scale (see the SI Section~A), the gap between $\s$ and $\sbvot$ is less than about $2\%$ (except for very few and typical recording errors of electoral data). Moreover, averaging $\s$ and $\sbvot$ over samples of $\approx200$ municipalities of similar sizes $N$ provides a difference less than $1\%$ for $N\gtrsim 1000$.  

In short, for large population municipalities, the bias introduced by the scale at which data are aggregated is weak and does not affect the main conclusions of the paper. \\ 

{\bf (2) Statistical effects due to large numbers}

Let us see if statistical fluctuations due to finite size effects considerably modify the expected values of involvement entropy. Indeed, For independent events, according to the central limit theorem (under conditions broadly applicable) fluctuations are on the order of $1/\sqrt{N}$. This is expected to be the case for the ratios $\pa$ and $\pn$, which should then lead to a bias in the entropy value. We want to estimate this bias and see if it is negligible (say less than $1\%$). To do so, we make a simulation with artificial data. For calibrating these data, we make use of the sample of the most populated municipalities. We measure the average values $\overline{\pa}$ and $\overline{\pn}$ of $\pa$ and $\pn$ over all municipalities in this sample of the largest municipality-size; and the corresponding standard deviations $\sigma_a$ and $\sigma_{bn}$. The surrogate data consists in a same number of ``municipalities", each one characterized by the same population size as in the empirical data. For these surrogate-municipalities, we draw the numbers of Abstentionists and of Null-Blank votes from binomial distributions, parametrized by the empirical average values and standard deviations of $\pa$ and $\pn$, as follows.

Let a surrogate-municipality with $N$ registered voters. Its numbers of Abstentionists, $\na$, and Null-Blank votes, $\nn$, are drawn from a binomial distribution such that:
\bea
& \na = {\cal B}(N\;;\;\overline{\pa}+\eta_a),\\
& \nn = {\cal B}(N\;;\;\overline{\pn}+\eta_{bn}),
\label{ebinom}
\eea
where $\eta_a$ and $\eta_{bn}$ are independent random Gaussian noises of mean $0$ and of standard deviation $\sigma_a$ and $\sigma_{bn}$, respectively. Note that here, for each citizen in a surrogate-municipality, probabilities to not vote and to put a null-blank vote are mutually independent.

Now, we can compare the average values $\smoy(N)$ of municipal involvement entropy in a sample of $\approx N$ surrogate-municipality-size, with $\smoy(N_{max})$ in the sample of the most populated municipalities. We find that the difference is less than $1\%$ when $N \gtrsim 2000$. In other words, for municipality-size greater than around $2000$, statistical fluctuations due to finite size effects are negligible for what concern the present study.\\

To conclude, we have seen that, for French electoral data, finite size effects do not affect significantly the municipal involvement entropy (i.e. by less than a $2\%$ deviation) for $N$ greater than $2000$. Note that $2000$ is much less than the typical municipality size of the most populated municipalities, for which the common value $\s\approx1$ is frequently found. Lastly, the same analysis done for other countries for which electoral data are also available at the polling station scale (see the SI Section~A) give the same results (see e.g. mean values of $\s$ over the 100 most populated municipalities, at the municipality scale in Tab.~\ref{si-tdata-mun}, compared to those at ballot box scale in Tab.~\ref{si-tdata-bvot}).\\

\begin{figure}[h!]
\centering
\includegraphics[width=8cm, height=6cm, clip=true]{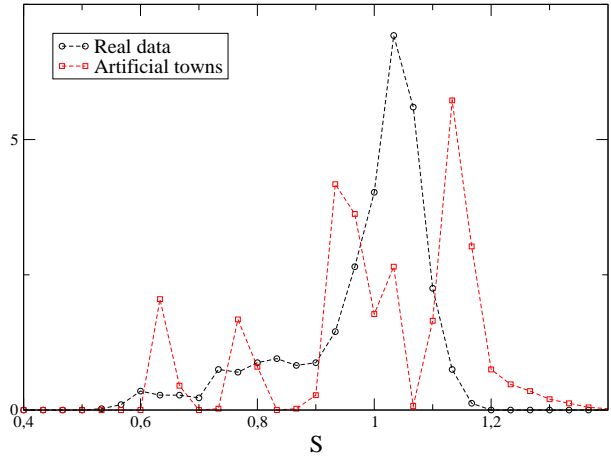}
\caption{{\bf Histograms of $\s$ of the 100 most populated towns compared with 100 artificial towns} (see text), in France, over elections since 2000.}
\label{fhisto-artificial}
\end{figure}

Now, let us show that the shape of the distribution of $\s$ over the 100 most populated towns (which is sharply peaked near $\s\approx 1$, apart from Austria) does not result from aggregating a large number of the citizen choices. In other words, the shape of the distribution of the involvement entropy for the 100 most populated towns (cf. Fig.~8-d) cannot be explained by a statistical bias due to a large number effect.

In order to see this point, 100 artificial town is created -- in France, without he loss of generality. Each artificial town results from the aggregation over 300 real small municipalities of real numbers of registered voters ($N$), abstentionists ($\na$), blank and null votes ($\nn$) and votes according to the list of choices ($\nc$). In other words, an artificial town comes from the aggregation of real citizen choices who live in small municipalities. Each municipality is taken into account only once. These 100 French artificial towns have artificial aggregated registered voters ($N$) from 7000 to 330000, and is equal to 34000 in average. Fig.~\ref{fhisto-artificial} allows one to compare the real distribution of $\s$ of the most populated French towns over elections since 2000 with the one which results from these 100 artificial towns. These two histograms are clearly different. 

To conclude, the shape of the distribution of the involvement entropy of most populated towns (cf. Fig.~8-d) is not due to a bias rooted in aggregating a large number of citizen choices. The shape itself depends on real citizen choices who live in these towns.

\clearpage
\subsection*{D. Logarithmic three choices value, $\mathbf{\tau_3}$, of polling stations}
As a supplement to the study of the entropy defined from the set of three ratios $\{\pa,\pc,\pn\}$, in this section we introduce another variable, called logarithmic three choices value, $\tau_3$, which also takes into account the set $\{\pa,\pc,\pn\}$. First, we show that the distribution of $\tau_3$, over polling stations in the 100 most populated municipalities appears stable over time, and also similar between different countries. Secondly, we justify our interest for this logarithmic three choices value from hypothesis on agents behavior. We compare two simple decision making rules, and give arguments against the most intuitive one, that of a choice decomposed into two successive binary choice decisions (first to vote or not, then to cast a valid vote or not). This confirms in a different way the existence of correlations between $\pa$ and $\pn$ (see Main text, Results Section, paragraph on ``Abstentions, valid votes and blank or null votes").

\subsubsection*{D.1. Logarithmic three choices value of polling stations in most populated towns}
\begin{figure*}[b!]
\centering
\includegraphics[width=16cm, height=4cm, clip=true]{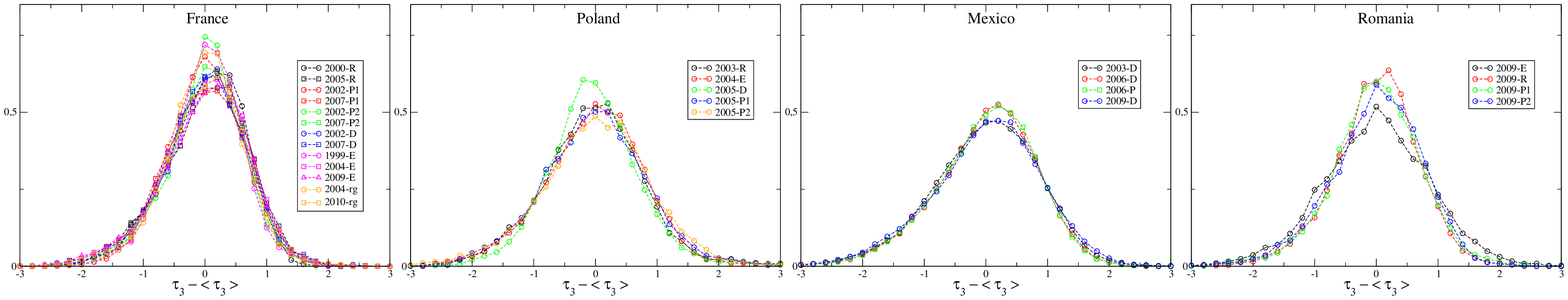}
\caption{{\bf Distribution over polling stations of the 100 most populated towns of $P(\tau_3 - \langle \tau \rangle)$ for each election}, where $\tau_3$ is the logarithmic three choices value and $\langle \tau_3 \rangle$ its average value over all concerned polling stations.}
\label{si-ftau3-histo}
\end{figure*}

In this section we generalize the analysis done in~\cite{si-these,si-diffusive_field}, where the statistics of the logarithmic turnout rate, $\tau=\ln \frac{\pa}{1-\pa}$, is studied. When considering the three possible values, $\pa, \pn, \pn$, the logarithmic three choices value $\tau_3$, as justified below Section D.2, can be defined by
\begin{equation}
\label{etau3bis} \tau_3 = \ln\big(\frac{\pc\cdot\pn}{(\pa)^2}\big). 
\end{equation}
Fig.~\ref{si-ftau3-histo} shows the pdf of the logarithmic three choices value $\tau_3$ over different polling stations of the 100 most populated towns in each country (apart from Canadian ones because more than third of polling stations have $\pn=0$, which lead to their logarithmic three choices values $\tau_3$ are undefined), i.e. the probability $P(\tau_3) {\rm d}\tau_3$ that a given polling station, inside the 100 most populated towns, has $\tau_3$ to within ${\rm d}\tau_3$. Although the average $\langle \tau_3 \rangle$ over these polling stations varies quite substantially between elections (see Fig.~\ref{si-ftau3-s} and Tab.~\ref{si-tdata-bvot}), the shape of the distribution of $\tau_3 - \langle \tau_3 \rangle$ is quite stable across elections for each country.

\begin{figure}[t!]
\begin{minipage}{0.5\linewidth}
\centering
\includegraphics[width=8cm, height=6cm, clip=true]{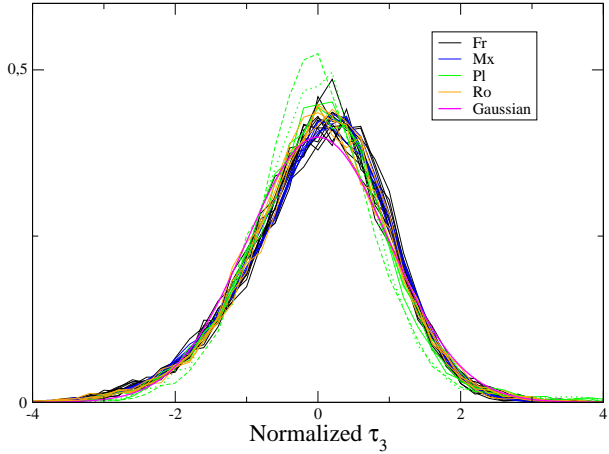}
\end{minipage}\hfill
\begin{minipage}{0.47\linewidth}
\caption{{\bf Distribution of normalized $\tau_3$ over polling stations of the 100 most populated towns} for 26 elections. The dotted line and the dashed line show respectively Pl-2003-R and Pl-2005-D elections. A normalized Gaussian is also plotted.}
\label{si-ftau3-norm}
\end{minipage}
\end{figure}

\begin{figure}[b!]
\begin{minipage}{0.5\linewidth}
\centering
\includegraphics[width=8cm, height=4cm, clip=true]{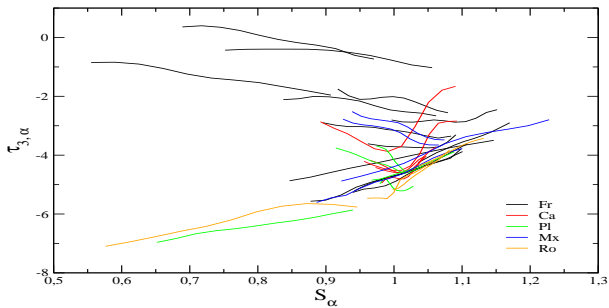}
\end{minipage}\hfill
\begin{minipage}{0.47\linewidth}
\caption{{\bf Logarithmic three choices value, $\tau_3$, with respect to involvement entropy $\s$.} Measures come from mean values of $\tau_3$ and $\s$ over polling stations in each of the 100 most populated towns. Curves are smoothed. Note that there is not a one-to-one relation between $\tau_3$ and $\s$.}
\label{si-ftau3-s}
\end{minipage}
\end{figure}

Consider now the normalized $\tau_3$ values, that is
$\hat{\tau_3}=\frac{\tau_3 - \langle \tau_3 \rangle}{\sigma}$, where $\langle \tau_3 \rangle$ and $\sigma$ are respectively the mean value and the standard deviation of $\tau_3$ over polling stations of the 100 most populated municipalities. Fig.~\ref{si-ftau3-norm} shows that the remarkable similarity between the distributions of these normalized logarithmic three choices for French, Mexican, Romanian and half Polish elections. A Kolmogorov-Smirnov test, where one only allows for a relative shift of the normalized distributions $P(\hat{\tau_3})$, over polling stations of the 100 most populated towns, does not allow one to reject the hypothesis that the distribution $P(\hat{\tau_3})$ is the same for all elections (except for half of the Polish elections).

Remark: there is not a one-to-one relation between the logarithmic three choices value, $\tau_3$, and the involvement entropy $\s$. Indeed, it is enough to invoke that the three ratios $\{\pa, \pc, \pn\}$ play a symmetric role for $\s$, and not for $\tau_3$. Fig.~\ref{si-ftau3-s} plots $\tau_3$ with respect to $\s$ for their average values over polling stations in each of the 100 most populated towns (see also Tab.~\ref{si-tdata-bvot} for basics statistics of $\s$ and $\tau_3$ over polling stations in the 100 most populated municipalities).

\subsubsection*{D.2. Towards a behavioral model in the three choices case}
Elaborating upon standard hypothesis on agents behavior, the goal of this section is to explain why $\tau_3$ defined above is the natural generalization of the logarithmic turnout rate introduced in~\cite{si-these,si-diffusive_field} for a single binary choice. \\

$\bullet$ {\bf Recall: Threshold decision rule for a single binary choice}\\
Let us first recall the rationale for introducing the logarithmic turnout rate. We consider $N$ agents making a binary decision. Agent $i$ makes its decision $n_i$, $n_i \in \{0,1\}$, according to 
\begin{equation}
n_i=\Theta(h_i+H),
\label{eq:bindec}
\end{equation}
with $\Theta(x \geq 0)=1$ and $\Theta(x < 0)=0$. Here $h_i$ is an idiosyncratic term characterizing the bias of agent $i$ in favor of the decision to vote ($n_i=1$). Idiosyncrasies are assumed independent random variables (hence uncorrelated between agents). $H$ is a global bias, a field identically applied to all agents, which can be seen as a `cultural field'~\cite{si-diffusive_field}. Note that here there is no direct interaction between agents.

According to this decision rule (Eq.~(\ref{eq:bindec})), in the large size limit $N\rightarrow \infty$, the fraction $\pv=1-\pa$ of decisions $1$ among the population, is equal to the cumulative distribution of idiosyncrasies: $\pv={\cal P}_>(-H)\equiv\int_{-H}^\infty P(h) {\rm d}h$. If idiosyncrasies are assumed to be distributed according to a logistic distribution~\cite{choice} of zero mean and of unity width, it comes that 
\bea
\label{etau}  \pv &=&\frac{1}{1+e^{-H}}\,,~~ \mathrm{or~ equivalently},\\ 
 H & = & \ln\big(\frac{\pv}{1-\pv}\big).
\eea
This justifies to study the statistics of the logarithmic turnout rate $\tau=\ln\big(\frac{\pv}{1-\pv}\big)$. As shown in~\cite{si-diffusive_field,si-turnout-stat}, the logarithmic turnout rate across French municipalities is remarkably stable over time, which allows one to make predictions that can be confronted with empirical observation~\cite{3-predictions,3-results}.\\

\begin{figure*}[h!]
  \centering
  \includegraphics[width=12cm, height=5cm, clip=true]{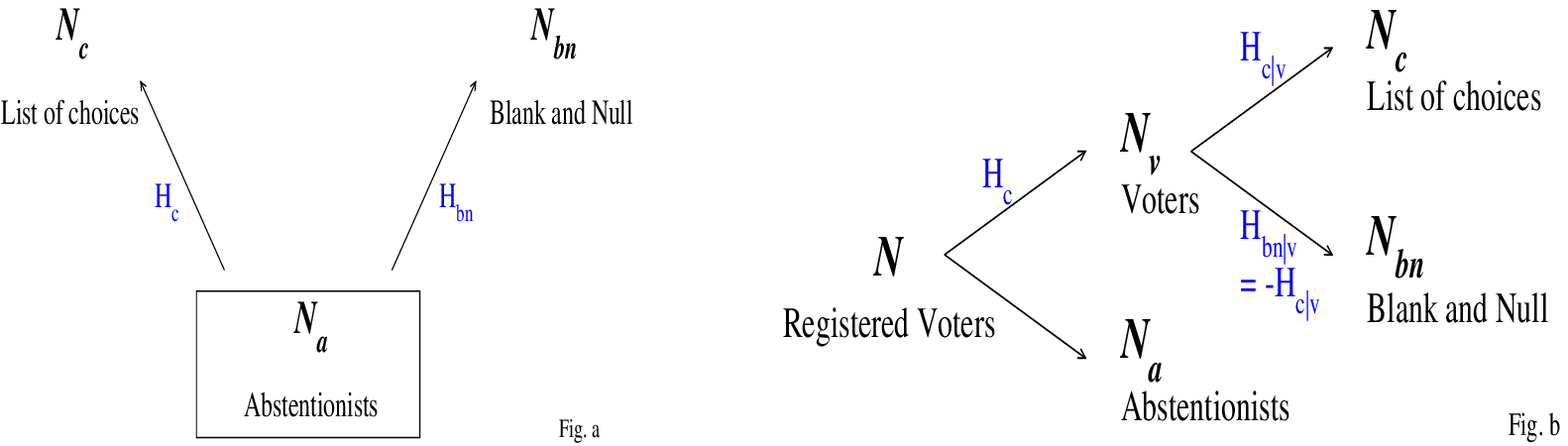}
  \caption{{\bf Two different hypothesis on agents behavior}: (a) two mutually exclusive binary decisions; (b) two sequential binary decisions.}
\label{si-ftau3-choice}
\end{figure*}

$\bullet$ {\bf Two binary decisions}\\
Now, we want to generalize to the case of three choices, not to vote, to cast a blank/null vote, and to cast a valid vote. One possibility is to assume a sequential decision (Figure \ref{si-ftau3-choice}, right): first to decide to vote or not, and if yes, then to decide to cast a blank/null vote, or to cast a valid vote. The alternative is to assume two mutually exclusive decisions (Figure \ref{si-ftau3-choice}, left). We explore both hypothesis, and show that data rule out the first one.

$\bullet$ {\bf Two mutually exclusive decisions}\\
Here we consider that, (1), to vote according to the list of choices, and (2), to cast a blank or null vote, are two mutually exclusive decisions. In other words, abstentionists are considered like a reservoir from which agents decide to make or not the choice (1) or the choice (2); moreover if they decide to do choice (1) (or conversely (2)), they do not decide anymore to make or not the choice (2) (or conversely (1)). Let $H_c$ the global field in favor of the choice (1) (to vote according to the list of choices), and $\pc^0$ the global ratio if choice (1) was unique, i.e. without any existence of choice (2) (see Fig.~\ref{si-ftau3-choice}-a). Conversely, $H_{bn}$ and $\pn^0$ refer to choice (2) (to put a blank or null vote) if it was a unique choice. From Eq.~(\ref{etau}), and again assuming a logistic distribution of idiosyncrasies, it comes
\bea \label{etau3-0}
& \pc^0 = {\cal P}_>(- H_c),\ &\mathrm{thus},\ H_c = \ln\big(\frac{\pc^0}{1-\pc^0}\big),\\
& \pn^0 = {\cal P}_>(- H_{bn}),\ &\mathrm{thus},\ H_{bn} = \ln\big(\frac{\pn^0}{1-\pn^0}\big).
\eea
Now choice (1) (respectively (2)), is made by agents who have not decided to make the other choice (2) (respectively (1)). Hence, the ratio $\pc$ (respectively $\pn$) is related to the ratio $\pc^0$ (respectively $\pn^0$) that would result from a single binary choice according to:
\bea \label{etau3-p-p0}
& \pc^0 = \frac{\pc}{1-\pn},\\
& \pn^0 = \frac{\pn}{1-\pc}.
\eea

Writing $H \equiv H_c+H_{bn}$, the sum of civic global fields applied to registered voters in this 3 choices process, Eqs.~(\ref{etau3-0},\ref{etau3-p-p0}) yield to~\footnote{When one of the three ratios $\{\pa, \pc, \pn\}$ is equal to zero, $\tau_3$ is undefined.}
\begin{equation}\label{etau3} H = \ln\big(\frac{\pc\cdot\pn}{(\pa)^2}\big). \end{equation}
Hence under the hypothesis of two mutually exclusive binary decisions, the quantity to study is the logarithmic three choices rate $\tau_3=\ln\big(\frac{\pc\cdot\pn}{(\pa)^2}\big)$. This is what is done above, Section D.1.\\

$\bullet$ {\bf Two sequential binary threshold decisions}\\
Now, let us consider the hypothesis of two sequential binary decisions.
The first binary decision is to vote or not to vote, and the second binary decision is to decide to cast a valid vote (according to the list of choices) or to put an invalid vote (i.e. a Blank or Null vote) given that the considered agent is a voter.

Let $H_v$ the global field related to the first decision, i.e. to vote (see Fig.~\ref{si-ftau3-choice}-b). The ratio of voters, $\pv$, over registered voters writes as:
\begin{equation} 
\label{ehv} \pv = {\cal P}_>(- H_v),\quad \mathrm{or},\quad H_v = \ln\big(\frac{\pv}{1-\pv}\big)\:.
\end{equation}
(Remind that $\pv=1-\pa=\pc+\pn$.) Let $H_{c|v}$ the global bias related to the second binary decision (given that the agent is a voter), that is the bias in favor to put a vote according to the list of choices. The ratio of votes according to the list of choice over voters is written as
\begin{equation}  
\label{ehcv} \frac{\pc}{\pv} = {\cal P}_>(- H_{c|v}),\quad \mathrm{or},\quad H_{c|v} = \ln\big(\frac{\frac{\pc}{\pv}}{1-\frac{\pc}{\pv}}\big)=\ln\big(\frac{\pc}{\pn}\big)\:.
\end{equation}
The second decision to put a Blank or Null vote is such that $H_{bn|v}=-H_{c|v}$ (since $\pn/\pv={\cal P}_>(- H_{bn|v})$ and $H_{bn|v}=\ln\big(\frac{\pn}{\pc}\big)$~).

According to this two sequential binary choices, the global field which leads a registered voter to put a Valid vote is $H'_c=H_v+H_{c|v}=\ln\big(\frac{\pv\cdot\pc}{\pa\cdot\pn}\big)$; and to put a Blank or Null vote is $H'_{bn}=H_v+H_{bn|v}=\ln\big(\frac{\pv\cdot\pn}{\pa\cdot\pc}\big)$. When Blank or Null ratio is very small ($\pn\ll 1$), $\pc\simeq\pa$, hence one has $H'_{bn}\simeq\tau_3$. So, statistics of $H'_{bn}$ are expected to be very similar to those of $\tau_3$.

If this sequential binary decisions point of view was correct, $H'_{bn}=\ln\big(\frac{\pv\cdot\pn}{\pa\cdot\pc}\big)$ and $H'_c=\ln\big(\frac{\pv\cdot\pc}{\pa\cdot\pn}\big)$ would share the same main features. However, this is strongly rejected by the empirical data. The shape of the distribution over polling stations of the 100 most populated towns of $H'_{bn} - \langle H'_{bn} \rangle$ is not constant from election to election (not shown here), this for each country over various elections. This is confirmed making use of Kolmogorov-Smirnov tests. \\

To conclude the analysis of the fluctuations around the mean of $\tau_3$, $H'_{bn}$ and $H'_c$ allows one to reject the hypothesis of a sequential binary choice, and to support the one of two mutually exclusive choices.

\clearpage
\subsection*{E. Looking for signs of tension, through polling stations analysis}

\begin{figure*}[h!]
\centering
\includegraphics[width=16cm, height=6cm, clip=true]{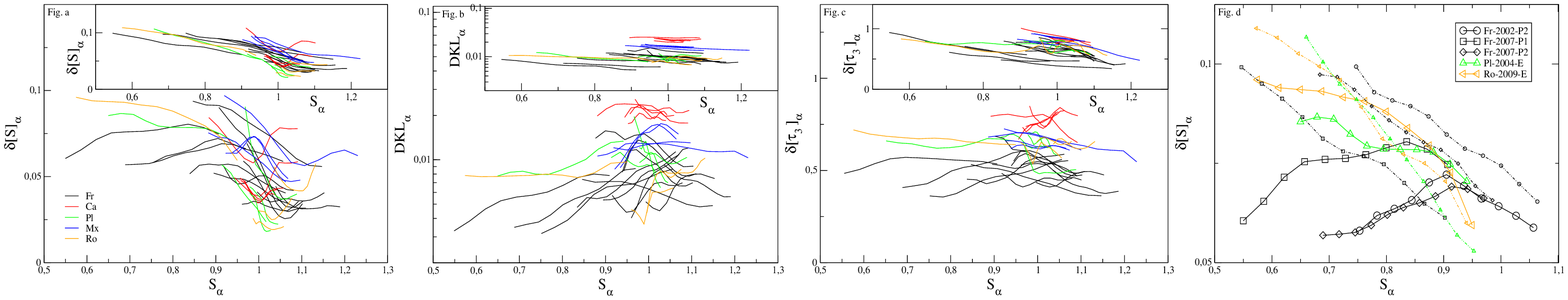}
\caption{{\bf Civic-involvement heterogeneity (at the polling station scale) in a town with respect to the involvement entropy of the town.} Curves are smoothed and concern the 100 most populated municipalities. Benchmark (see text) curves are plotted in the insets. Heterogeneity measures result from standard deviation of involvement entropy of polling stations (\ref{si-ftension}-a), Kullback–Leibler divergence between a polling stations and other polling stations of the town (\ref{si-ftension}-b), and standard deviation of logarithmic 3 choices value of polling stations (\ref{si-ftension}-c). Fig.~\ref{si-ftension}-d: same as Fig.~\ref{si-ftension}-a, but restricted for the 5 elections which deviate the more from $\s\approx1$, where plain lines and dashed line plot respectively real data and benchmark curves.}
\label{si-ftension}
\end{figure*}

This section seeks to detect some `tension', in connection with the involvement entropy. We follow the assumption that `tension' have some effects for polling stations heterogeneity inside a town. In other words, we try to detect some specific variation of polling stations heterogeneity in a given town, in connection with the involvement entropy of this town. Polling stations (inside a same town) heterogeneity is investigated by three different ways: (1) standard deviation of involvement entropies over all polling stations of the considered town; (2) Kullback–Leibler divergence from one polling station compared to other polling station of the town; (3) standard deviation of the logarithmic three choices value (because the shape of its distribution is stable, see the SI Section~D) over all polling stations of the town.

The analysis uses polling stations inside the 100 most populated towns (see the SI Section~A for more details). Real results will be compared to a benchmark. The benchmark is based on the same heterogeneity of ratios $\pa$ (idem for $\pc$, and $\pn$) of polling stations for every town.\\

Let a town and a polling station of this town respectively called $\alpha$ and $\alpha_i$. The polling station $\alpha_i$ has some measures, for instance its number of registered voters $N_{\alpha_i}$, and the set of 3 ratios $\{\paai, \pcai, \pnai\}$ that provides its involvement entropy $\sai$ and its logarithmic three choices value $\tai$. The average over all the polling stations of the town (weighted by the number of registered voters, $N_{\alpha_i}$), gives the corresponding value for the whole town $\alpha$, e.g. the set of 3 ratios $\{\paa, \pca, \pna\}$, its involvement entropy $\sa$, and its logarithmic three choices value $\ta$. The weighted (by the number of registered voters) standard deviation over all the polling stations $\alpha_i$ of the town $\alpha$ is written as $\delta[..]_\alpha$, like for instance $\delta[\pa]_\alpha$, etc., $\delta[\s]_\alpha$ and $\delta[\tau_3]_\alpha$. These quantify the heterogeneities within the town $\alpha$.

Fig.~\ref{si-ftension}-a plots involvement entropy heterogeneity of a town $\alpha$ (i.e. $\delta[\s]_\alpha$) with respect to its involvement entropy (i.e. $\sa$). One should pay attention to the fact that $\delta[\s]_\alpha$ going trough a minimum as $\sa\approx1$ could just be a consequence of $|{\rm d}S|$ having a minimum near $\pn\approx0$ and $\pa\approx0.5$. Hence the benchmark presented here consists in comparing the empirical data with surrogate ones for which the heterogeneity in $\pa$, $\pc$ an $\pn$ is the same for all municipalities, up to a binomial noise.

Here, the benchmark forces the same heterogeneity of the set of ratios $\{\pa, \pc, \pn\}$ for every town, but keep their initial value of $\{\pa, \pc, \pn\}$ for the whole town. In other words, let a town $\alpha$, $\delta[\pa]_\alpha$, $\delta[\pc]_\alpha$ and $\delta[\pn]_\alpha$ have the same values than in other towns; but $\{\paa, \pca, \pna\}$ are the real values of the town $\alpha$, measured by the election. 

The benchmark is realized as follows. First, we measure for each town $\alpha$, $\paa$ and $\pna$; and also $\delta[\pa]_\alpha$ and $\delta[\pn]_\alpha$. The average values of heterogeneities $\delta[\pa]_\alpha$ and $\delta[\pn]_\alpha$ over the 100 considered towns are respectively written as $\overline{\delta_{p_a}}$ and $\overline{\delta_{p_{bn}}}$. Secondly, we drawn from a binomial distribution, for each polling station $\alpha_i$, its number of registered voters who do not take part to the election ($N_{a,\,\alpha_i}$) and the number of Blank and Null votes ($N_{bn,\,\alpha_i}$), such that:
\bea
& N_{a,\,\alpha_i} = {\cal B}(N_{\alpha_i}\;;\;\paa+\eta_a),\\
& N_{bn,\,\alpha_i} = {\cal B}(N_{\alpha_i}\;;\;\pna+\eta_{bn}),
\label{ebenchmark}
\eea
where $\eta_a$ and $\eta_{bn}$ are independent Gaussian noises of mean $0$ and of standard deviation $\overline{\delta_{p_a}}$ and $\overline{\delta_{p_{bn}}}$ respectively, and $N_{\alpha_i}$ is the real number of registered voters of the polling station $\alpha_i$ of the considered town $\alpha$. Note that we use a binomial distribution in order to take into account finite size effects; and here, for each citizen in a surrogate-polling station, probabilities to not vote and to put a null-blank vote are mutually independent.

Instead of making use of standard-errors, an alternative measure of heterogeneity is provided by making use of the so-called Kullback–Leibler divergence which characterizes the difference between two probability distributions. For each polling station ${\alpha_i}$ of a given town $\alpha$, we compute the divergence $DKL_{\alpha_i}$ from the polling station distribution $P_{\alpha_i}$ to the rest of the town, $Q_{\alpha_i} \equiv P_{\alpha - \alpha_i}$,
\begin{equation}
DKL_{\alpha_i} \equiv \sum_j P_{\alpha_i}(j)\log \frac{P_{\alpha_i}(j)}{Q_{\alpha_i}(j)}
\end{equation}
where, here and in the following, for any distribution we write $P(j), j=1,2,3$, instead of $\pa, \pc, \pn$. Then we compute the mean Kullback–Leibler divergence, $DKL_\alpha$, of the town $\alpha$ by averaging over all polling stations, weighting by the corresponding number of registered voters, $N_{\alpha_i}$. 
\begin{equation}
DKL_{\alpha} \equiv \sum_i \frac{N_{\alpha_i}}{N_{\alpha}}\; DKL_{\alpha_i}
\end{equation}
This mean Kullback–Leibler divergence $DKL_\alpha$ gives us a measure of heterogeneity of polling stations into a town $\alpha$.\\

Fig.~\ref{si-ftension} compares benchmarks curves and empirical data~\footnote{In this section, extreme values greater than \textit{3 sigma} are note taken into account in order to remove some electoral errors, etc.}. It appears that, the smaller the involvement entropy $\s$ (with $\s\lesssim0.85$), the smaller the involvement entropy heterogeneity at the polling station level (see more specifically Fig.~\ref{si-ftension}-d). Heterogeneity of polling stations in a same town is measured via three different ways: standard deviations of the involvement entropy and the logarithmic three choices ratio, and also via the Kullback–Leibler divergence. In other words, the more the town is ``ordered" (for its electorate civic-involvement), the more the town is homogeneous (at the polling station scale, and still for a civic involvement point of view). Note also that this point is particularly clear when the ratio $\pc$ is high (e.g. for 3 French elections), compared to cases where $\pa$ are high (e.g. for European Parliament elections in Romania and Poland). It can also be noted that often heterogeneity of involvement entropies of polling stations inside towns ($\delta[\s]_\alpha$) has a significantly minimal value when their involvement entropies ($\sa$) are around $1$, and this minimization is much more marked for real data than for benchmark ones.

\clearpage
\subsection*{F. Disentangling Blank votes from Null votes}

\begin{figure*}[h!]
\centering
\includegraphics[width=16cm, height=7.5cm, clip=true]{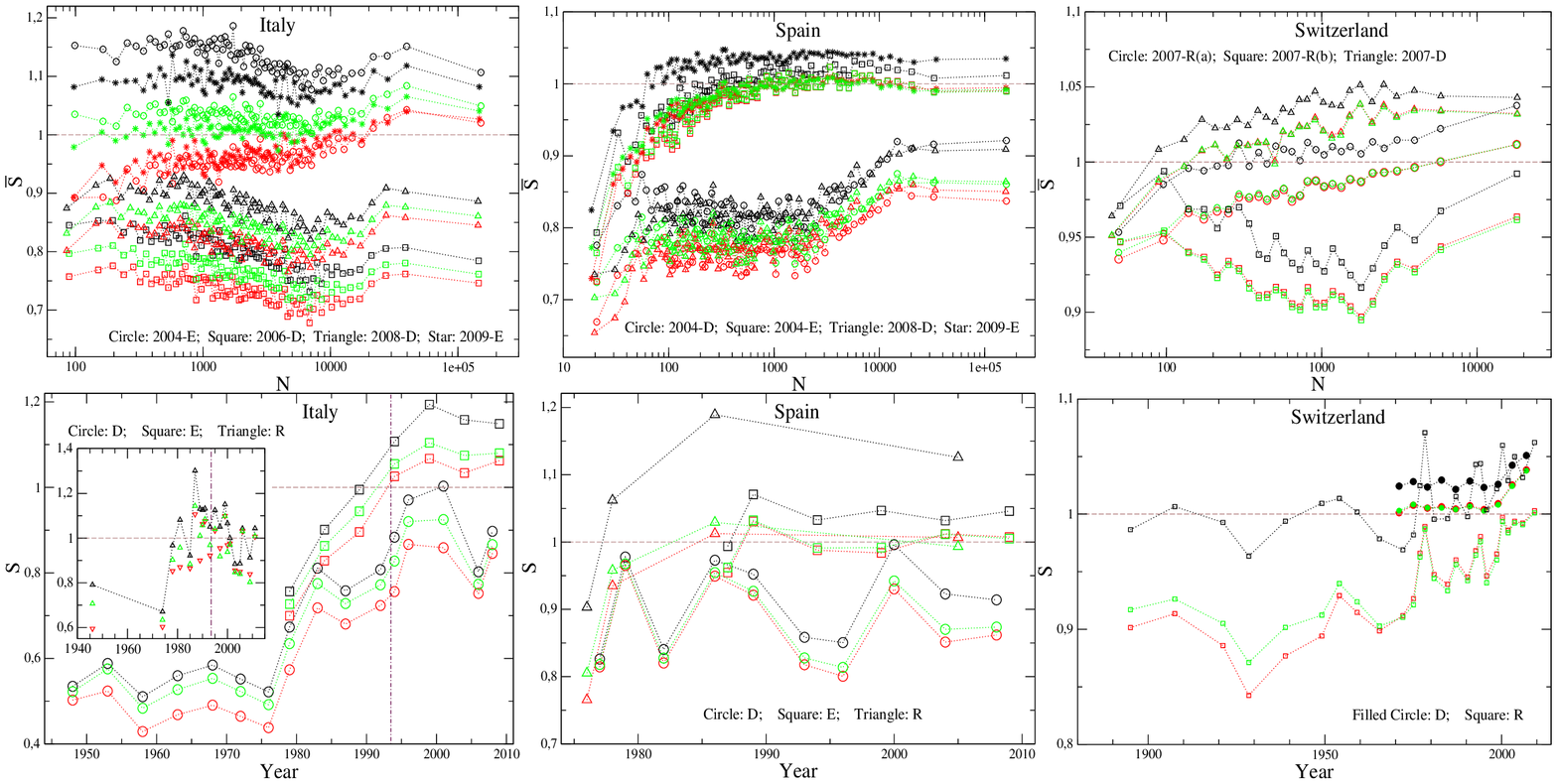}
\caption{Blank votes are grouped with: (1) Null votes (like in the main text, cf. Eq.~(2)) in black; (2) Valid Votes or another vote included in the list of choices (cf. Eq.~(\ref{es3-wc})) in red; (3) citizens who do not take part to the election (cf. Eq.~(\ref{es3-wa})) in green. Top: mean values of $\s$, $\swc$, $\swa$, over bins with around 100 municipalities of size $\approx N$ (like in Fig.~5). Bottom: Evolution in time of $\s$, $\swc$, $\swa$ (with the same scale of aggregate data as in Figs.~6 and \ref{si-fdyn}). For the sake of clarity, standard deviations over Swiss {\it Cantons} and Spanish {\it Comunidades aut\'onomas} are note shown. Each point (R) for Swiss graph gives the average of around 20 Swiss referendums. The end of Italian compulsory voting is shown by a vertical line.}
\label{si-fblancs}
\end{figure*}

Italy~\footnote{We only analyze the first question asked in a Referendum. Senate elections are note shown in Fig.~\ref{si-fblancs}-below because they are very similar to Chamber of Deputies (D) elections.}, Spain and Switzerland~\footnote{Chamber of deputies elections (D) distinguish, in our database, Blank vote between Null votes since 1971; and since 1887 for {\it votations} (or referendums).} are countries for which Blank votes are distinguished from Null votes. Let $\nwh$ and $\nnu$ the number of citizens who respectively vote Blank and Null amongst $N$ registered voters (of one municipality, {\it Canton}, {\it Comunidad aut\'onoma}, the whole country). Ratios, or probabilities, to respectively vote Blank and Null are
\begin{equation}\label{ep-blancs} \pwh = \frac{\nwh}{N},\quad \pnu = \frac{\nnu}{N}.\end{equation}

In such cases, it is legitimate to consider that Blank votes should be categorized with votes in favor of one of a the proposed choices to the election. Then, the Blank vote has not a `marginal' involvement meaning, like previously, but its citizen involvement is similar to another Valid vote according to the list of choices of the election. One should then consider a modified involvement entropy, defined from the 3-set ratios (of sum unity) $\{\pa, (\pc+\pwh), \pnu\}$, that is
\begin{equation}\label{es3-wc} \swc = -\pa \log(\pa)-(\pc+\pwh) \log(\pc+\pwh)-\pnu\log(\pnu).\end{equation}

Alternatively, one may consider that Blank votes loose their `marginal' aspect in citizen involvement, and should be categorized as votes from citizen who do not take part to the election. Then the relevant modified involvement entropy, defined from the 3-set ratios (still of sum unity) $\{(\pa+\pwh), \pc, \pnu\}$, writes as  
\begin{equation}\label{es3-wa} \swa = -(\pa+\pwh)\log(\pa+\pwh)- \pc\log(\pc)-\pnu\log(\pnu).\end{equation}

Figures~\ref{si-fblancs} shows for Italy, Spain and Switzerland, the involvement entropy, $\s$, and the modified versions, $\swc$ and $\swa$: (1) for municipalities and with respect to the municipality-size $N$ (as in Fig.~5); (2) for the whole country (directly for Italy, and as a weighted mean by population-size over 25 or 26 Swiss {\it Cantons} and 17 or 19 Spanish {\it Comunidades aut\'onomas}) as a function of time (as in Figs.~6 and \ref{si-fdyn}). Fig.~13 shows the modified involvement entropy $\swc$ ($\swa$ which is not shown, is very close to $\swc$), with respect to the involvement entropy $\s$, for $\sim 530$ Swiss Referendums.\\

Figure~\ref{si-fblancs} exhibits some trends and regularities that depend on the values of involvement entropy $\s$. (1) When $\s < 1$ (e.g. in Italian and Spanish Chamber of Deputies elections, both at municipality scale or at large scale of aggregate data), modified involvement entropies are smaller than $\s$. This means a greater order of the modified citizen involvement. It can be interpreted as follows: the loss of nuance or specificity (for citizen involvement) that Blank vote have, implies a greater polarization or heterogeneity of the electorate, still split into 3 groups. (2) When $\s\approx1$, two different cases arise. First, for Spanish European Parliament elections, Swiss Chamber of Deputies elections and Referendums (uniquely for the latter, since the 2000s), both at municipality scale or at large scale of aggregate data: the modified involvement entropies are slightly lower than $\s$, but still $\approx1$. Second, for earlier Swiss Referendums, and particularly before the 1960s: $\swc$ (or $\swa$) are lower than $\s$, but not slightly lower. (3) When $\s > 1$ and $\s \ncong1$ (e.g. for Italian European Parliament elections, both at municipality scale or at large scale of aggregate data, and Spanish Referendums, particularly 1986 and 2005 ones, at provincial scale), modified involvement entropies are still lower than $\s$. But one more time, it is surprising to notice that modified involvement entropies are such that $\swc \approx 1$ (or $\swa \approx 1$). It can be explained as follows: subtlety or specificity of citizen involvement due to Blank votes means an increasing of disorder of the electorate involvement. The loss of this subtlety or specificity (i.e. considering Blank votes like another vote in favor of the list of choices, or like another abstentionist) implies a loss of `tension' contained in electoral campaign. And strikingly, this loss of `tension' provides a new entropy, like the usual common-value of involvement entropy, $\approx1$.

Note that above items (1) and (3) (i.e. when significantly $\s<1$ or $\s>1$), pointed out in Fig.~\ref{si-fblancs}, are clearly shown in Fig.~13 for Swiss Referendums. (In the Fig.~13, $\s'$ means $\swc$, which is very near to $\swa$ on average.) Note also that the surprising plateau (which provides modified involvement entropies equal to $\approx 1$, on average, when $\s\gtrsim 1.05$) does not exist, in our database, for most populated municipalities. For the latter case (not shown), the around 100 most populated municipalities for which $\s\gtrsim 1.05$, uniquely provides $\swc$ (or $\swa$) lower than $\s$ such that, on average, $\overline{\swc}\approx1$ (or $\overline{\swa}\approx1$), but without a plateau.

Lastly, this study does not allow us to know whether it is more meaningful (according to the entropy of the electorate involvement) to consider Blank vote like another vote proposed in the list of choices or like another abstentionist vote. Nevertheless, in our database, Blank votes seem more meaningful that Null Votes in Spain and in Switzerland. Indeed, when $\pnu$ and $\pwh$ are interchanged between each other in Eqs.~(\ref{es3-wc}) or (\ref{es3-wa}), above item (3) (when significantly $\s > 1$, then the modified involvement entropy is $\approx1$) does clearly not exist for Spanish and Swiss Referendums.\\

To conclude, let us recall the main point of this section: when involvement entropy $\s$ does not obey to the common occurrence $\s\approx1$ for high population-size municipalities, or at large aggregate scale, because the citizen involvement of the electorate is too much disordered (i.e. significantly $\s>1$), then the modified involvement entropy (by the loss of the specificity of Blank votes) takes on average the same common value $\approx1$.

\clearpage

\end{document}